\pdfoutput=1
\RequirePackage{ifpdf}
\ifpdf %We are running pdfTeX in pdf mode
\documentclass[pdftex]{sigma}
\else
\documentclass{sigma}
\fi

\newtheorem{Theorem}{Theorem}[section]

\newtheorem{Lemma}[Theorem]{Lemma}
\newtheorem{Proposition}[Theorem]{Proposition}

\DeclareMathOperator{\const}{const}

\numberwithin{equation}{section}

\begin{document}

\allowdisplaybreaks

\newcommand{\arXivNumber}{1404.7234}

\renewcommand{\PaperNumber}{114}

\FirstPageHeading

\ShortArticleName{Periodic Vortex Streets and Complex Monodromy}

\ArticleName{Periodic Vortex Streets and Complex Monodromy}

\Author{Adrian D.~HEMERY~$^\dag$ and Alexander P.~VESELOV~$^{\ddag\S}$}

\AuthorNameForHeading{A.D.~Hemery and A.P.~Veselov}

\Address{$^\dag$~Charterhouse School, Godalming, Surrey, GU7 2DX, UK}
\EmailD{\href{mailto:adh@charterhouse.org.uk}{adh@charterhouse.org.uk}}

\Address{$^\ddag$~Department of Mathematical Sciences, Loughborough University,\\
\hphantom{$^\ddag$}~Loughborough, Leicestershire, LE11 3TU, UK}
\EmailD{\href{mailto:A.P.Veselov@lboro.ac.uk}{A.P.Veselov@lboro.ac.uk}}
\Address{$^\S$~Moscow State University, Russia}

\ArticleDates{Received August 28, 2014, in f\/inal form December 10, 2014; Published online December 23, 2014}

\Abstract{The explicit constructions of periodic and doubly periodic vortex relative equilibria using the theory of
monodromy-free Schr\"odinger operators are described.
Several concrete examples with the qualitative analysis of the corresponding travelling vortex streets are given.}


\Classification{76B47; 34M05; 81R12}

\section{Introduction}

The study of vortex dynamics is a~classical subject going back to Helmholtz~\cite{Helm}.
If we identify the plane with the set of complex numbers $\mathbb C$ then the dynamics of~$N$ point vortices
$z_1(t),\dots, z_N(t)$ with circulations (or, vorticities) $\Gamma_1, \dots, \Gamma_N$ is determined by the system
\begin{gather*}
\frac{d\bar z_j}{dt} = \frac{1}{2\pi i} \sum\limits_{k \neq j}^{N} \frac{\Gamma_k}{z_j-z_k},
\qquad
j=1, \dots, N.
\end{gather*}
In the periodic setting we have the equations
\begin{gather*}%\label{per}
\frac{d\bar z_j}{dt} = \frac{1}{2\pi i} \sum\limits_{k \neq j}^{N} \Gamma_k \cot(z_j-z_k),
\qquad
j=1, \dots, N,
\end{gather*}
where we assume for simplicity that the period $L=\pi$ (see~\cite{AS1, MST} and the pioneering paper by Friedman and
Polubarinova~\cite{FP}).

In this paper we consider the {\it periodic relative equilibria} of the vortices described by the system
\begin{gather}
\label{pereq}
\frac{1}{2\pi i}\sum\limits_{k \neq j}^{N} \Gamma_k \cot(z_j-z_k)-\bar v=0,
\qquad
j=1, \dots, N,
\end{gather}
where $v= \frac{dz_j}{dt}$ is the common constant velocity of the vortices.
A~classical example is given by the so-called {\it von K\'arm\'an vortex street}~\cite{Lamb, Kar,Kar2}, corresponding to the
case $N=2$, $\Gamma_1+\Gamma_2=0$.

\begin{figure}[h] \centering

\includegraphics[width=8cm]{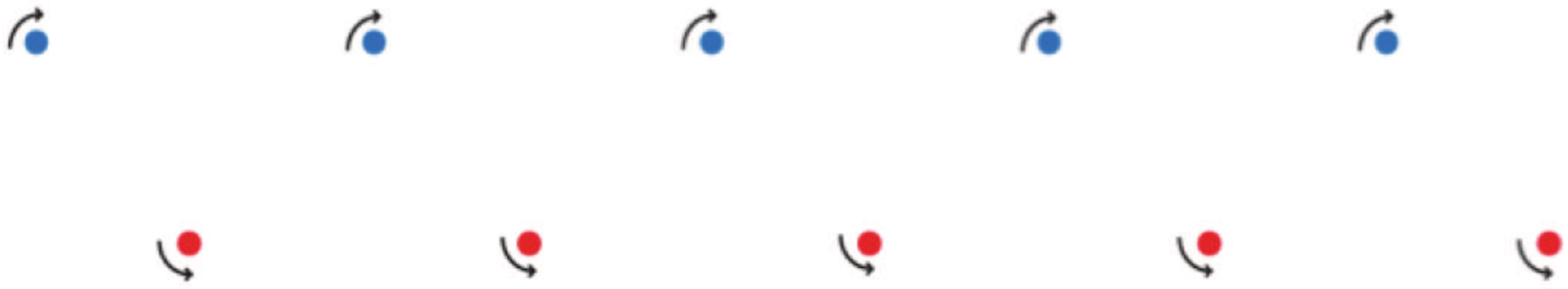}

\caption{Classical von K\'arm\'an vortex street moving to the left.}\label{KarmanX2}
\end{figure}

More recently the vortex dynamics in periodic domains was studied in~\cite{AS1, MST, Strem, Tok}, but the dynamics in
general is known to be non-integrable and even the description of all relative equilibria remains a~largely open
question.

One of the results of this paper is the following class of the explicit periodic relative vortex equilibria, most of
which seem to be new.

Let $k=(k_1,\dots, k_n)$, $k_1< k_2 <\dots < k_n$ be a~set of distinct natural numbers, and $\phi=(\phi_1,\dots,\phi_n)$,
$\phi_j \in \mathbb C/\pi \mathbb Z$ be a~set of~$n$ complex numbers considered modulo $\pi$. Let
\begin{gather}
\label{W1}
W_{k, \phi}(z) =W\big(\chi_{k_1,\phi_1}(z), \dots, \chi_{k_n,\phi_n}(z)\big)
\end{gather}
be the Wronskian with $\chi_{k_j,\phi_j}(z)=\sin (k_j z + \phi_j)$ and
\begin{gather}
\label{W2}
W_{k, \phi, \kappa}(z) =W\big(\chi_{k_1,\phi_1}(z), \dots, \chi_{k_n,\phi_n}(z), e^{i\kappa z}\big),
\end{gather}
where $\kappa \in \mathbb C$ is another complex parameter.
Consider the combined conf\/iguration $\Sigma_{k,\phi, \kappa}$ of complex roots of $W_{k, \phi, \kappa}(z)=0$ and $W_{k,\phi}(z)=0$.
Prescribe their circulations as follows: a~zero~$z_i$ of~$W_{k, \phi, \kappa}(z)=0$ has the circulation $\Gamma_i=m_i$
equal to the multiplicity of~$z_i$ while for a~zero~$z_j$ of~$W_{k, \phi}(z)$ the circulation $\Gamma_j=-m_j$ is
negative multiplicity of $z_j$ (for common zeros of $W_{k, \phi, \kappa}(z)$ and $W_{k, \phi}(z)$ the circulation is the
dif\/ference of the corresponding multiplicities).

\begin{Theorem}\label{Theorem1}
The configuration $\Sigma_{k,\phi, \kappa}$ is a~periodic relative vortex equilibrium moving with constant velocity
$v=-\frac{\bar \kappa}{2\pi}$ for any non-critical $\kappa \notin \{k_1,\dots, k_n\}$.
In the frame moving with the vortices the complex potential of the flow is
\begin{gather*}
W= \frac{1}{2\pi i} \log \psi (\kappa,z),
\end{gather*}
where
\begin{gather*}
%\label{psi}
\psi(\kappa, z) = \frac{W_{k, \phi, \kappa}(z)}{W_{k, \phi}(z)}
\end{gather*}
is a~trigonometric Baker--Akhiezer function
for the corresponding monodromy-free Schr\"odinger operator
\begin{gather}
\label{poten}
L= -D^2+ \sum \frac{m_j(m_j-1)}{\sin^2(z-z_j)},
\qquad
D=\frac{d}{dz}.
\end{gather}
For critical value $\kappa=k_j$ we have an equilibrium vortex configuration with complex potential $W= \frac{1}{2\pi i}
\log \psi_j(z)$,
\begin{gather*}%\label{psicrit}
\psi_j = \frac{W_{k^{(j)}, \phi^{(j)}}(z)}{W_{k, \phi}(z)},
\end{gather*}
where $k^{(j)}$, $\phi^{(j)}$ are the sets~$k$, $\phi$ without $k_j$ and $\phi_j$ respectively.
\end{Theorem}

The proof (see Section~\ref{Section3} below) is based on a~simple observation that the conditions of relative equilibrium coincide
with the {\it Stieltjes relations}~\cite{V} and thus hold for all periodic trigonometric monodromy-free operators of the
form~\eqref{poten}.
Such operators play an important role in the theory of Huygens principle as it was shown by Berest and
Loutsenko~\cite{BL}.
They were classif\/ied in~\cite{B, CFV} (see Theorem 4.3 in~\cite{CFV}) and all turned out to be iterated Darboux
transformations applied to trivial potential $u=0$.
For rational potentials a~similar observation was known already for quite a~while~\cite{Bartman} (see also~\cite{ANSTV}
and references therein), but in the periodic setting we have not seen this in the literature although it looks quite
natural.

\begin{figure}[h] \centering \includegraphics[width=8cm]{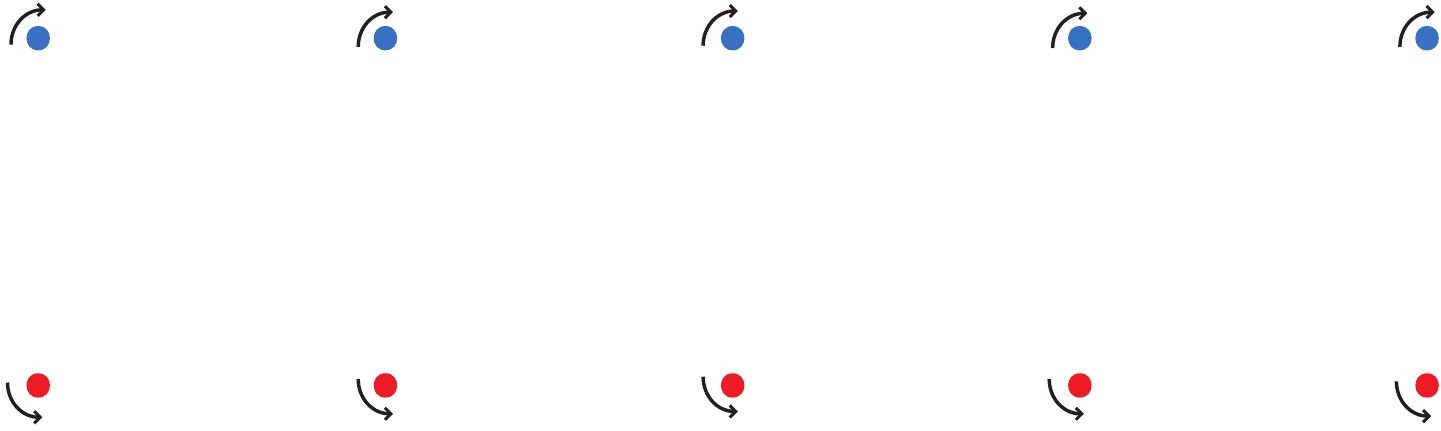}
\caption{Fast von K\'arm\'an vortex street, known to be unstable.}
\label{Karman}
\end{figure}

Von K\'arm\'an vortex streets correspond to the simplest case $n=1$. Indeed, let us consider for simplicity $k_1=1$, $\phi_1=0$, then
\begin{gather*}
\psi(\kappa,z)=(i\kappa - \cot z)e^{i\kappa z}.
\end{gather*}
We have one zero and one pole modulo~$\pi$: the pole is $z=0$ and the zero is the solution of $\cot z=i\kappa$, which is
equivalent to
\begin{gather*}
e^{2iz}=\frac{\kappa+1}{\kappa-1}
\end{gather*}
if $\kappa\neq 1$. Assuming that~$\kappa$ is real and positive, we have 2 cases: $\kappa>1$ (fast), $\kappa<1$ (slow)
when
\begin{gather*}
z=\frac{1}{2i}\log \frac{\kappa+1}{\kappa-1}
\qquad
\text{and}
\qquad
z=\frac{1}{2i}\log \frac{\kappa+1}{1-\kappa} +\frac{\pi}{2}
\end{gather*}
respectively.
In the slow case we have the von K\'arm\'an street shown in Fig.~\ref{KarmanX2}, the fast case is shown in
Fig.~\ref{Karman}.

The complex potential of the f\/low in the moving frame is $W=\frac{1}{2\pi i} \log \psi(\kappa,z)$, the instantaneous
complex potential of the f\/low in the f\/ixed frame is
\begin{gather*}
W=\frac{1}{2\pi i} \log \psi(\kappa,z)e^{-i\kappa z}=\frac{1}{2\pi i} \log (i \kappa -\cot z).
\end{gather*}
For the critical value $\kappa=1$ the zero of~$\psi$ goes to inf\/inity: $\psi=\frac{1}{\sin z}$ and we have trivial
vortex equilibrium, consisting of points $l\pi$, $l \in \mathbb Z$ with circulations $-1$.

\begin{figure}[h] \centerline{\includegraphics[width=6.5cm]{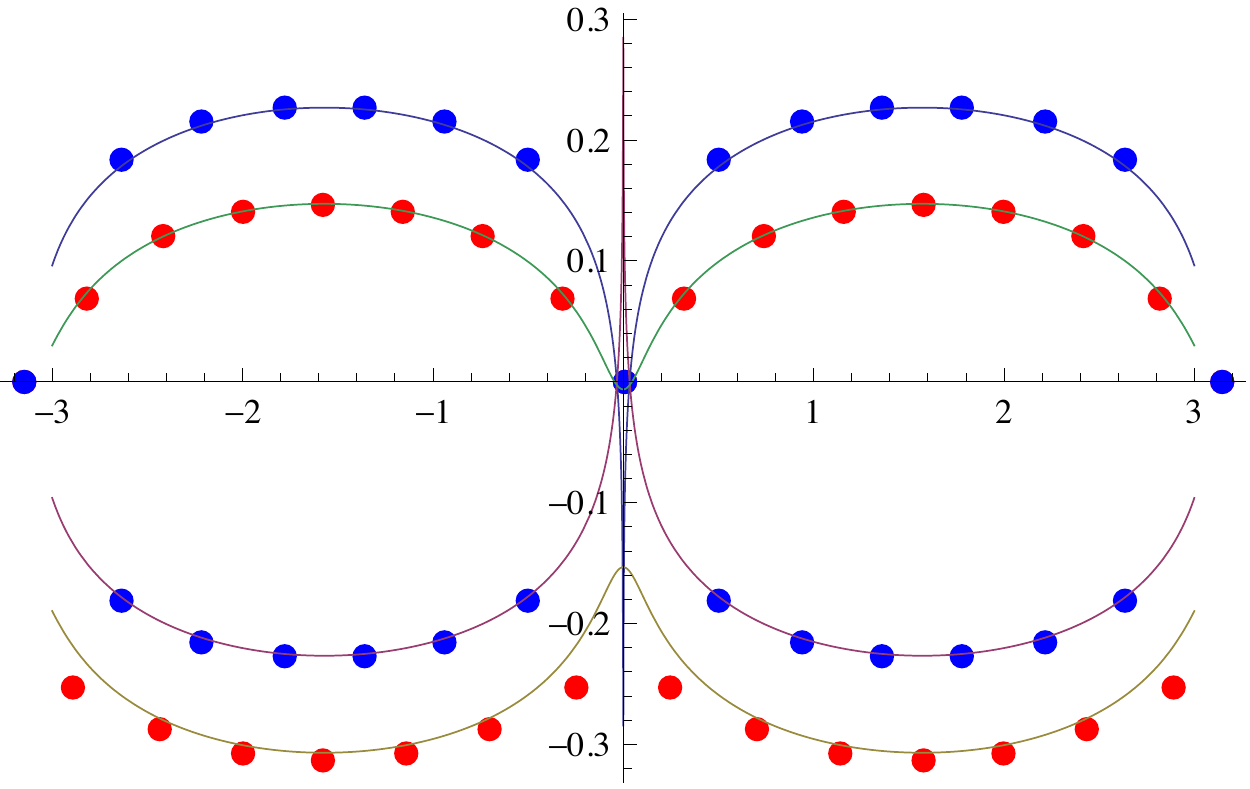} \hspace{2pt} \includegraphics[width=6.5cm]{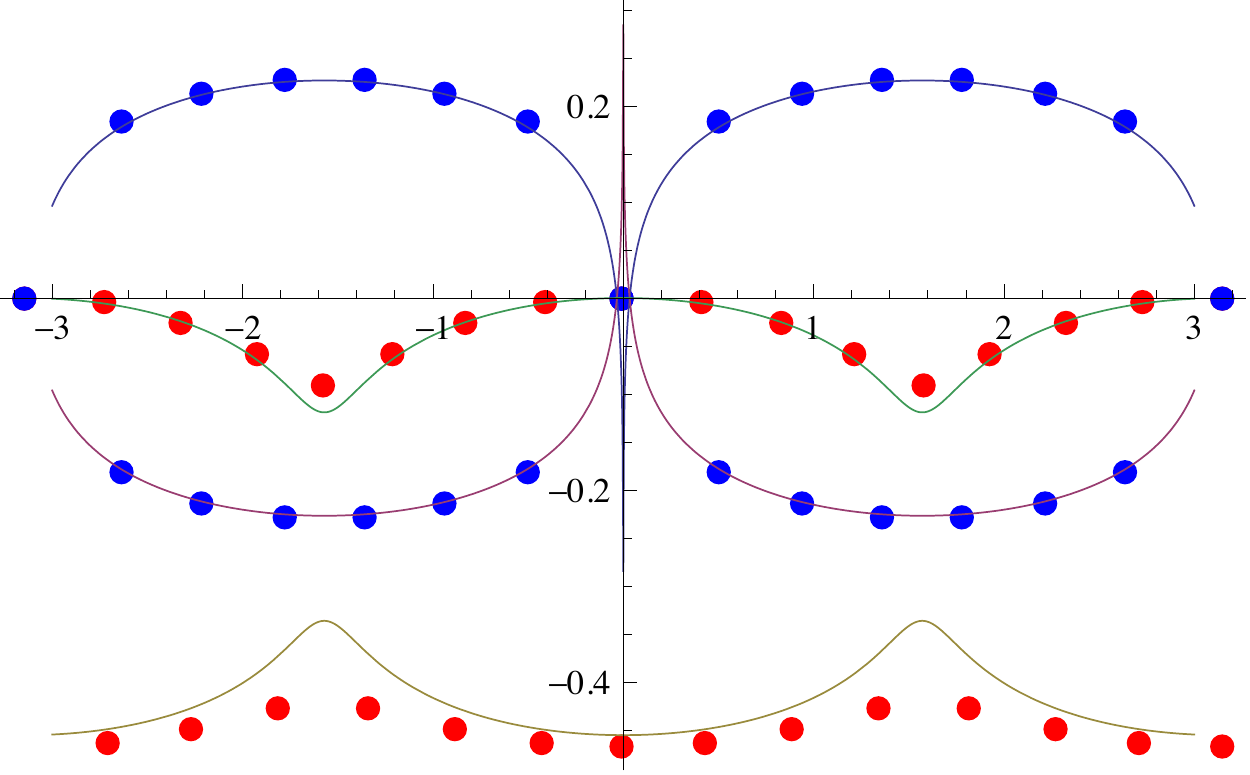}}
\caption{Vortex streets $\Sigma_{k,0,\kappa}$ with $k=(7,8)$, $\kappa=4$ (left) and $\kappa=7.4$ (right) with the graphs
of the asymptotic curves~\eqref{triglimcurve},~\eqref{movingcurve}.}
\label{WKIntro}
\end{figure}

Two examples in the case $n=2$ are shown in Fig.~\ref{WKIntro}, the details and more examples with some qualitative analysis are presented in Section~\ref{Section6}.
All the pictures in the paper were produced using Mathematica.
The colour of the points indicates the sign of the circulations (which are generically $\pm 1$): red means positive,
blue~-- negative circulations.
The axes on all f\/igures are~$x$ and~$y$, such that $z=x+iy$.

Our approach based on the ideas of~\cite{V} is very general and can be applied to all monodromy-free operators.
In particular, we apply it to the classical Whittaker--Hill operator and its Darboux transformations~\cite{HV} to
construct periodic equilibria in the presence of background f\/low with complex potential $W=A \cos 2z$.

In the doubly periodic case we have more general result, when only one meromorphic solution is required.
Let $\sigma(z)$ and $\wp(z)$ be classical Weierstrass elliptic functions~\cite{WW}.

\begin{Theorem}\label{Theorem2}
Let
\begin{gather}
\label{psi}
\psi(z)=\prod\limits_{i=1}^{N} \sigma^{m_i}(z-z_i) e^{Bz}
\end{gather}
with $m_i \in \mathbb Z$, $m_1+\dots+m_N=0$ and $B\in \mathbb C$ be a~solution of the Schr\"odinger equation
\begin{gather}
\label{eq}
-\psi''+\left(\sum\limits_{i=1}^N m_i(m_i-1)\wp(z-z_i)\right)\psi=E\psi.
\end{gather}
Then $W(z)=\frac{1}{2\pi i}\log \psi(z)$ is the complex potential of a~doubly periodic relative vortex equilibrium in
the moving frame.

Conversely, let the set $z_1, \dots, z_N$ with integer circulations $m_1, \dots, m_N$ with zero sum be a~doubly periodic
relative vortex equilibrium.
Then the function~\eqref{psi} with suitable constant~$B$ is a~solution of equation~\eqref{eq} for some energy~$E$.
\end{Theorem}

The classical theory of Lam\'e equation going back to Hermite~\cite{WW} and modern theory of elliptic
solitons~\cite{AMM, DN, Krichever, S, T, Take, TV, V2}, which can be def\/ined as the theory of monodromy-free operators
of the form~\eqref{eq} (or Picard potentials in terminology of~\cite{GW}), provides many examples of such solutions and
thus new doubly periodic relative equilibria.
In particular, the logarithm of the corresponding elliptic Baker--Akhiezer function is a~complex potential for a~relative
doubly periodic vortex equilibrium in the moving frame.
We should mention though that in contrast to trigonometric case an ef\/fective description of all elliptic f\/inite-gap (or
algebro-geometric) operators still remains an open problem, which was f\/irst emphasized by S.P.~Novikov as part of the ef\/fectivisation programme in f\/inite-gap theory.

Note that in all the examples we produce the circulations are integers, which might be useful for applications to liquid
helium, where circulations are known to be quantized~\cite{Feynman, Tkac}.

\section{Monodromy-free Schr\"odinger operators and Stieltjes relations}

Consider the Schr\"odinger operator
\begin{gather*}
L=-D^2+u(z)
\end{gather*}
in the complex domain $z \in \mathbb C$ with meromorphic potential $u(z)$ having poles only of second order.
The operator~$L$ is called {\it monodromy-free} if the corresponding Schr\"odinger equation{\samepage
\begin{gather}
\label{ode}
-\varphi''+u(z)\varphi = E \varphi
\end{gather}
has all solutions meromorphic for all values of $E$.}

Near a~pole (which can be assumed for simplicity to be $z = 0$) the potential can be represented as Laurent series $u =
\sum\limits_{i = -2}^{\infty} c_{i} z^{i}$. Following the classical Frobenius approach one can look for the solutions of
the form
\begin{gather*}
\varphi = z^{-\mu}\left(1 + \sum\limits_{i=1}^{\infty} \xi_i z^{i}\right).
\end{gather*}
The corresponding~$\mu$ must satisfy the characteristic equation $\mu(\mu + 1) = c_{-2}$, which means that the
equation~\eqref{ode} has a~meromorphic solution only if the coef\/f\/icient $c_{-2}$ at any pole has a~very special form:
\begin{gather}
\label{loc1}
c_{-2} = m(m+1),
\qquad
m\in {\bf Z}_+.
\end{gather}
This condition is in fact not suf\/f\/icient: the corresponding solution~$\varphi$ may have a~logarithmic term.
The following important lemma due to Duistermaat and Gr\"unbaum~\cite{DG} gives the conditions when this does not
happen.

\begin{Lemma}[\protect{\cite{DG}}]
The logarithmic terms are absent for all~$\lambda$ if and only if in addition to~\eqref{loc1} all the first $m+1$ odd
coefficients at the Laurent expansion of the potential are vanishing:
\begin{gather*}%\label{loc2}
c_{2k-1} = 0,
\qquad
k = 0, 1,\dots, m.
\end{gather*}
\end{Lemma}

Let $\psi(z)$ be a~solution of the corresponding Schr\"odinger equation
\begin{gather*}
\big(-D^2+u(z)\big)\psi(z)=\lambda \psi(z)
\end{gather*}
and $f(z) = D \log \psi(z)$. Then the potential $u(z)$ can be expressed as
\begin{gather*}
u(z)=f'+f^2+\lambda.
\end{gather*}

\begin{Proposition}[\protect{\cite{V}}]
\label{Proposition1}
Let~$f$ be a~meromorphic function having the poles of the first order with integer residues.
The Schr\"odinger operator~$L$ with the potential $u = f' + f^2+\const$ is monodromy-free if and only if at any pole
$z_0$ with ${\rm Res}_{z = z_0}\, f = m$
the following generalised Stieltjes relations are satisfied:
\begin{gather*}
{\rm Res}_{z = z_0}\, f^2 = {\rm Res}_{z = z_0}\, f^4 =\dots= {\rm Res}_{z = z_0}\, f^{2|m|} = 0.
\end{gather*}
\end{Proposition}

The proof is simple: substituting
\begin{gather*}
f = \frac{\pm m}{z-z_0} + \sum\limits_{k=0}\alpha_k (z-z_0)^k
\end{gather*}
with $m\in {\mathbb Z}_+$ into $u = f' + f^2$ that $c_{-2} = m(m \pm 1)$ we can check that the trivial monodromy
conditions $c_{2k-1} = 0$, $k = 0, 1,\dots, m-1$ are equivalent to the vanishing of the coef\/f\/icients $\alpha_{2k} = 0$, $k =0, 1,\dots, m-1$.
The last relation $c_{2m-1} = 0$ is then fulf\/illed automatically, see~\cite{V}.

In particular, we always have the original Stieltjes relation: at every pole $z_i$ of~$f$
\begin{gather*}
{\rm Res}_{z = z_i}\, f^2 =0,
\end{gather*}
which Stieltjes used to give electrostatic interpretation of the zeroes of some classical polyno\-mials~\cite{S1}.
We are going to use the same idea to produce some new relative vortex equilibria.

\section{Trigonometric monodromy-free operators\\ and periodic vortex streets}\label{Section3}

As we have already mentioned all~$\pi$-periodic trigonometric monodromy-free operators of the form~\eqref{poten} are
known~\cite{B, CFV} to be the result of several Darboux transformations applied to $L_0 = -D^2$.

The corresponding potentials have the form
\begin{gather}
\label{pot}
u(z)=-2D^2 \log W\big(\chi_{k_1,\phi_1}(z), \dots, \chi_{k_n,\phi_n}(z)\big),
\end{gather}
where $k_1<k_2<\dots <k_n$ are distinct natural numbers, $\phi_i \in \mathbb C/\pi \mathbb Z$ are arbitrary complex
numbers modulo $\pi$,
\begin{gather*}
\chi_{l,\phi}(z)=\sin (l z + \phi),
\qquad
l \in \mathbb N.
\end{gather*}
The Schr\"odinger equation
\begin{gather*}
\big({-}D^2+u(z)\big)\psi=\kappa^2 \psi
\end{gather*}
with $u(z)$ given by~\eqref{pot} has solution
\begin{gather*}
\psi(\kappa,z)=\frac{W_n(\kappa,z)}{W_n(z)},
\end{gather*}
where $W_n(z)=W_{k, \phi}(z)$, $W_n(\kappa, z)=W_{k, \phi, \kappa}(z)$ are the Wronskians~\eqref{W1},~\eqref{W2}.
Note that $W_n(\kappa, z)=P_n(\kappa, z)e^{i\kappa z}$, where $P_n(\kappa, z)=W_n(z) i^n \kappa^n +
\sum\limits_{j=0}^{n-1} A_j(z) \kappa^j$ is a~polynomial in~$\kappa$ with coef\/f\/icients being trigonometric polynomials
in $z$, so the ratio $\psi(\kappa,z)=W_n(\kappa,z)/W_n(z)$ is (a version of) the corresponding {\it trigonometric
Baker--Akhiezer function}~\cite{Kri}.

Let
\begin{gather*}
W_n(z)=C \prod\limits_{i=1}^M \sin^{m_i} (z-a_i),
\qquad
W_{n}(\kappa, z)=C' \prod\limits_{i=1}^N \sin^{n_i}
(z-b_i)e^{i\kappa z}
\end{gather*}
with some constants~$C$ and $C'$ be the corresponding factorisations with possible multiplicities.
Then the log-derivative $f=D \log \psi(\kappa,z)=D \log W_{n}(\kappa,z)-D \log W_{n}(z)$ has the form
\begin{gather*}
f=\sum\limits_{i=1}^M m_i \cot (z-b_i)- \sum\limits_{j=1}^N n_j \cot (z-a_j)+i\kappa=\sum\limits_{i=1}^{M+N} \Gamma_i \cot (z-z_i)+i\kappa,
\end{gather*}
where $z_1,\dots, z_{M+N} = a_1, \dots, a_M, b_1, \dots, b_N$ and $\Gamma_i=m_i$ for $i=1,\dots, M$ and
$\Gamma_{M+j}=-n_j$ for $j=1, \dots,N$. It may happen that $a_i=b_j$, so we have a~cancelation with $\Gamma=m_i-n_j$ in
that case.

We claim that the set of vortices with position at $z_1,\dots, z_{M+N}$ with the corresponding circulations $\Gamma_1,
\dots, \Gamma_{M+N}$ described above is a~periodic relative vortex equilibrium conf\/iguration moving with velocity
$v=\bar \kappa/2\pi$.
Indeed, by Stieltjes relations
\begin{gather*}
{\rm Res}_{z = z_j}\, f^2=2 \Gamma_j\sum\limits_{k\neq j} \Gamma_k \cot (z_j-z_k)+2i\Gamma_j \kappa=0
\end{gather*}
for all $j=1, \dots, M+N$. Comparing with~\eqref{pereq} we see that{\samepage
\begin{gather*}
\frac{d\bar z_j}{dt} = \frac{1}{2\pi i} \sum\limits_{k \neq j} \Gamma_k \cot(z_j-z_k)=-\frac{\kappa}{2\pi},
\end{gather*}
so we have the vortex conf\/iguration moving with constant speed $v=-\bar \kappa/2\pi$.}

Note that in the frame moving with the vortices the corresponding f\/low can be written as
\begin{gather*}
\frac{d\bar z}{dt} = \frac{1}{2\pi i} \left(\sum\limits_{i=1}^M m_i \cot (z-b_i)- \sum\limits_{j=1}^N n_j \cot (z-a_j)+i\kappa\right)
= \frac{d}{dz} \frac{1}{2\pi i} \log \psi,
\end{gather*}
so by def\/inition $W=\frac{1}{2\pi i} \log \psi(\kappa, z)$ is the {\it complex potential} of the f\/low~\cite{Acheson,
M-T}\footnote{There is a~sign discrepancy in the def\/inition of the complex potential of the f\/low between
Milne-Thomson~\cite{M-T} and Acheson~\cite{Acheson}.
We follow Acheson here.}.
The instantaneous complex potential of the f\/low in the f\/ixed frame is
\begin{gather*}
%\label{inst}
W_*=W-\frac{\kappa}{2\pi} z =\frac{1}{2\pi i} \log \big(\psi e^{-i\kappa z}\big) = \frac{1}{2\pi i} \log
\frac{P_n(\kappa,z)}{W_n(z)}.
\end{gather*}

At the critical level $\kappa=k_j$ the Wronskian $W_{k, \phi, \kappa}(z)$ reduces to
\begin{gather*}
W_{k, \phi, k_j}(z)=W_{k^{(j)}, \phi^{(j)}}(z),
\end{gather*}
so we have
\begin{gather*}
\psi=\frac{W_{k^{(j)}, \phi^{(j)}}(z)}{W_{k, \phi}(z)},
\end{gather*}
where $k^{(j)}=(k_1, \dots, \hat k_j, \dots, k_n)$, where $\hat k_j$ means that $k_j$ is omitted from the list, and
similarly for $\phi^{(j)}$. Some of the zeros disappear at inf\/inity and the relative equilibriium becomes a~genuine one,
in agreement with the general claim by Montaldi, Soli\`ere and Tokieda~\cite{MST} that if the sum of the vorticities is
not zero the only relative equilibria are the usual equilibria.
This completes the proof of Theorem~\ref{Theorem1}.

We will discuss many examples of corresponding relative vortex equilibria in Section~\ref{Section6}.
Here we will mention only new {\it collinear vortex equilibria} (when all the vortices are on a~real line), related to
Baker--Akhiezer conf\/igurations found by M.~Feigin and D.~Johnston~\cite{FJ12}.

The corresponding vortex equilibrium $\Sigma(n,m,l)$ depends on integer parameters $n > m\geq 1$ and even $l>0$.
It corresponds to
\begin{gather*}
k=(1,2,\dots, n-m, n-m+2, n-m+4, \dots, n+m-2, n+m+l)
\end{gather*}
(or, more explicitly $k_j=j$ for $j=1,\dots,n-m$, $k_{n-m+j}=n-m+2j$ for $j=1,\dots, m-1$ and $k_n=n+m+l$) and $\phi=0$.
One can show that in this case
\begin{gather*}
W_{k^{(n)}, \phi^{(n)}}(z)=C_1 \cos^{\frac{m(m-1)}{2}}z \sin^{\frac{n(n-1)}{2}}z,
\\
W_{k, \phi}(z)=C_2 \cos^{\frac{m(m+1)}{2}}z \sin^{\frac{n(n+1)}{2}}z \prod\limits_{j=1}^l \sin(z - z_j),
\end{gather*}
where $z_1,\dots, z_l \in (0,\pi)$ are certain simple real zeros located symmetrically with respect to $\pi/2$
(see~\cite{FJ12}).
This means that the function $W=\frac{1}{2\pi i} \log \psi(z)$ with
\begin{gather*}%\label{newFJ}
\psi(z)=W_{k, \phi}(z)/W_{k^{(n)}, \phi^{(n)}}(z)=C \cos^{m}z \sin^{n}z \prod\limits_{j=1}^l \sin(z - z_j)
\end{gather*}
is the complex potential of the collinear vortex equilibrium conf\/iguration $\Sigma(n,m,l)$ consisting of $z=0$ with
circulation~$n$, $z=\pi/2$ with circulation~$m$ and $z=z_1,\dots,z_l$ with circulation~1.

One can produce more equilibriums by changing $z \rightarrow qz$ and extend this conf\/iguration to a~moving vortex street
by considering $\psi(z)=W_{k, \phi, \kappa}(z)/W_{k, \phi}(z)$ with non-critical~$\kappa$.

\section{Whittaker--Hill equation and relative vortex equilibria\\ in the background f\/low}

Since the Darboux transformation preserves the class of monodromy-free operators one can use it in a~more general
situation to produce vortex equilibria in the presence of a~background f\/low.

We demonstrate it here in the example of the classical Whittaker--Hill equation
\begin{gather*}%\label{W-H}
-\psi'' -\big(4\alpha s \cos2x + 2\alpha^2\cos4x\big)\psi = \lambda \psi.
\end{gather*}
This equation is special because for natural values of parameter~$s$ it has precisely~$s$ elementary eigenfunctions of
the form $\psi_j(z)=\varphi_j(z)e^{\alpha \cos 2x}$, $j=1,\dots, s$, where $\varphi_j(z)$ are some trigonometric
polynomials~\cite{W}.
For example, for $s=3$ we have
\begin{gather*}
\varphi_1=1 + \frac{\sqrt{1+16\alpha^2}-1}{4\alpha}\cos2x,
\qquad
\varphi_2=\sin 2x,
\qquad
\varphi_3=\frac{1 - \sqrt{1+16\alpha^2}}{4\alpha} + \cos2x.
\end{gather*}

Let $I=\{i_1,\dots, i_n\}$ be a~set of distinct natural numbers,
\begin{gather*}
W_I=W(\psi_{i_1}, \dots, \psi_{i_n})=W(\varphi_{i_1}, \dots, \varphi_{i_n})e^{n\alpha\cos 2x}
\end{gather*}
be the Wronskian of the corresponding eigenfunctions.
Following~\cite{HV} consider the Darboux transformation of the Whittaker--Hill operator with the potential
\begin{gather*}
\tilde u = -\big(4\alpha s \cos2x + 2\alpha^2\cos4x\big) -2D^2 \log W_I,
\end{gather*}
and its eigenfunction $\psi_{JI}(z)=W_J(z)/W_I(z)$, where $J=\{i_1,\dots, i_n, i_{n+1}\}$,
\begin{gather*}
W_J=W(\psi_{i_1}, \dots, \psi_{i_n}, \psi_{i_{n+1}})=W(\varphi_{i_1}, \dots, \varphi_{i_n},
\varphi_{i_{n+1}})e^{(n+1)\alpha\cos 2x}.
\end{gather*}
The corresponding log-derivative $f=D \log \psi_{JI}(z)$ has the form
\begin{gather*}
f=\sum\limits_{i=1}^M m_i\cot(z-b_i)-\sum\limits_{j=1}^N n_j\cot(z-a_j)-2\alpha\sin 2z=\sum\limits_{i=1}^{M+N}\Gamma_i\cot(z-z_i)-2\alpha\sin2z,
\end{gather*}
where as before $a_i$ and $b_j$ are the zeros of the denominator $W_I$ and numerator $W_J$ with circulations being
negative multiplicities and multiplicities respectively.

\begin{figure}[h] \centering \includegraphics[width=6cm]{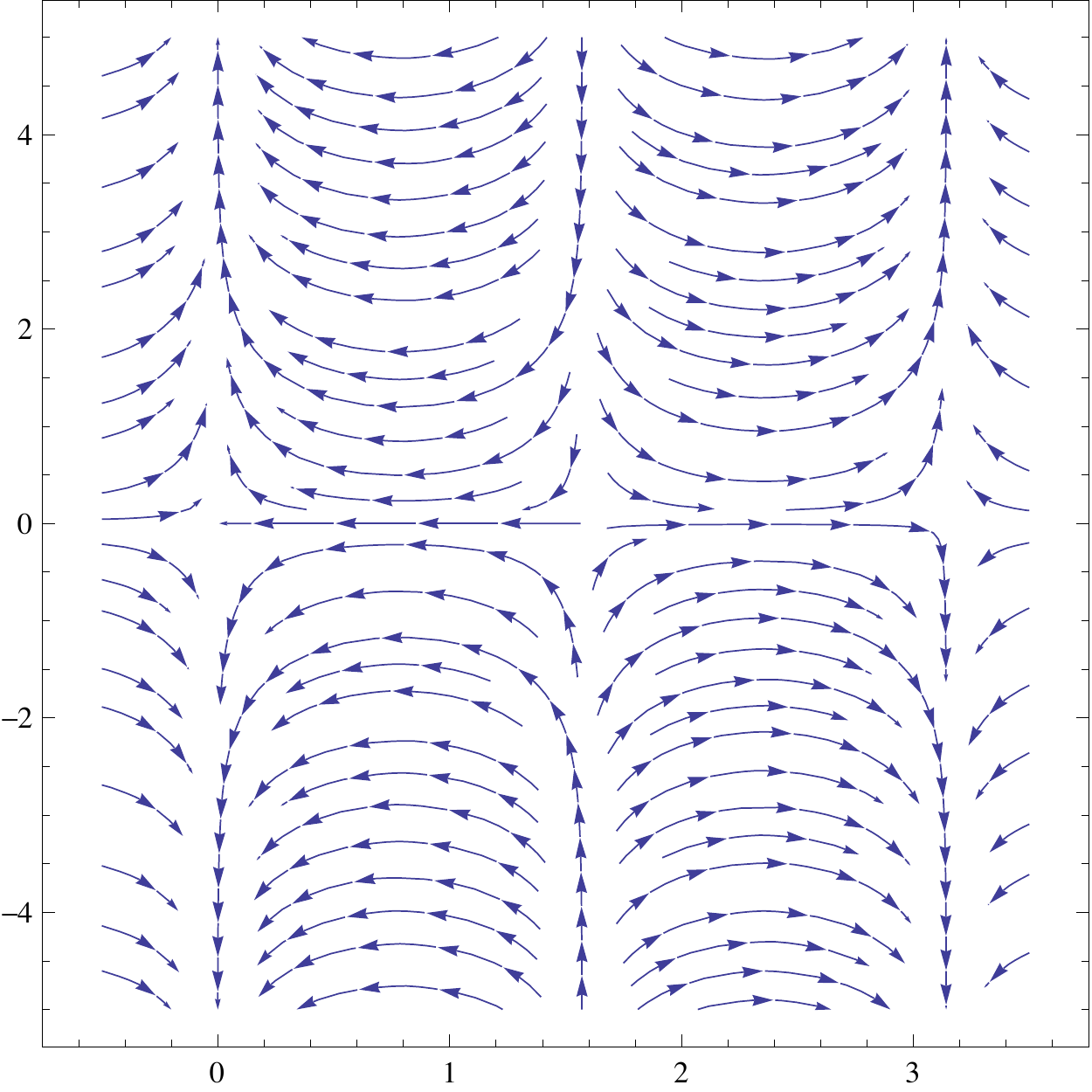}
\caption{The background f\/low with complex potential
$W=\cos 2z$.}
\label{WHBack}
\end{figure}

By Proposition~\ref{Proposition1} we have
\begin{gather*}
{\rm Res}_{z = z_i}\, f^2=2 \Gamma_i \sum\limits_{j\neq i} \Gamma_j \cot (z_i-z_j)-4 \Gamma_i\alpha \sin 2z_i=0,
\end{gather*}
or,
\begin{gather*}
\sum\limits_{j\neq i} \Gamma_j \cot (z_i-z_j)-2\alpha \sin 2z_i=0
\end{gather*}
for all $i=1, \dots, M+N$. These Stieltjes relations can be interpreted as the vortex equilibrium relations in
a~periodic background f\/low (see Fig.~\ref{WHBack}).

\begin{Theorem}
The function $W=\frac{1}{2\pi i} \log \psi_{JI}(z)=\frac{1}{2\pi i}(\log W_J(z) - \log W_I(z))$ is the complex potential
of a~vortex equilibrium in the background flow with complex potential $\frac{\alpha}{2\pi i} \cos 2z$.
\end{Theorem}

\begin{figure}[h] \centering \includegraphics[width=6cm]{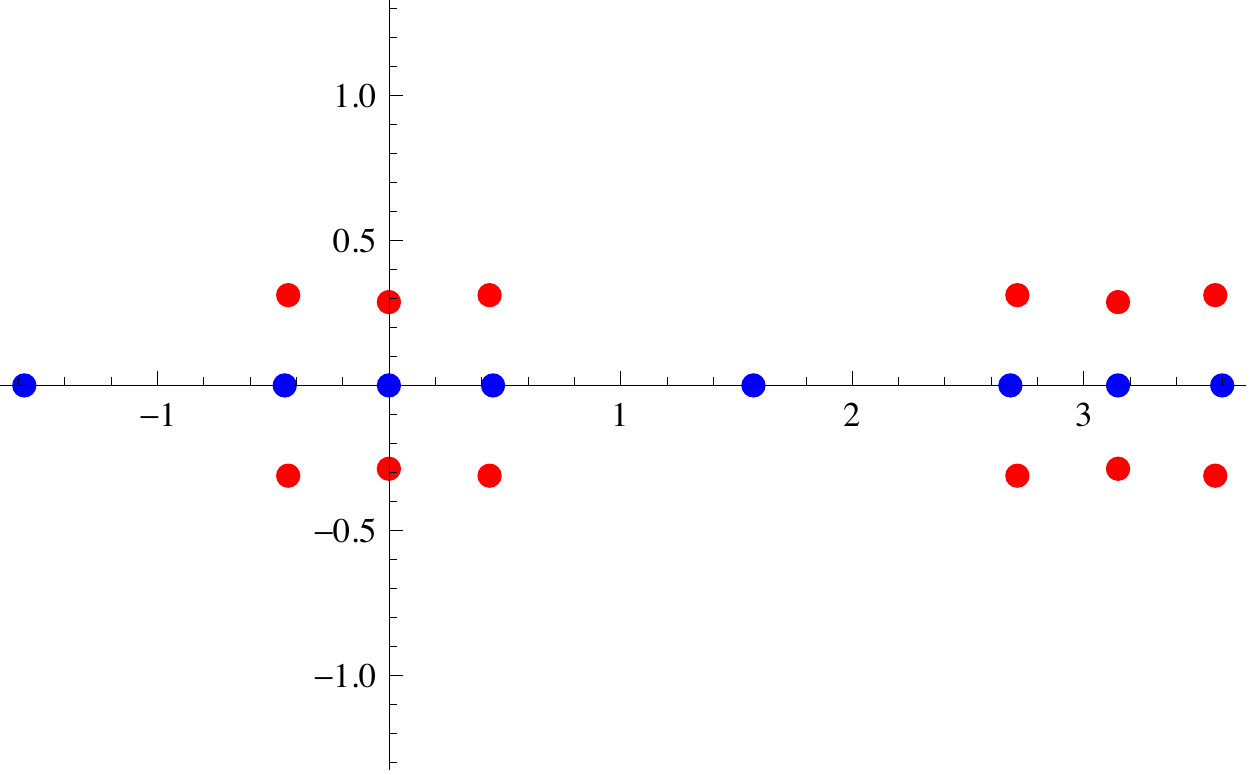} \hspace{5pt} \includegraphics[width=6cm]{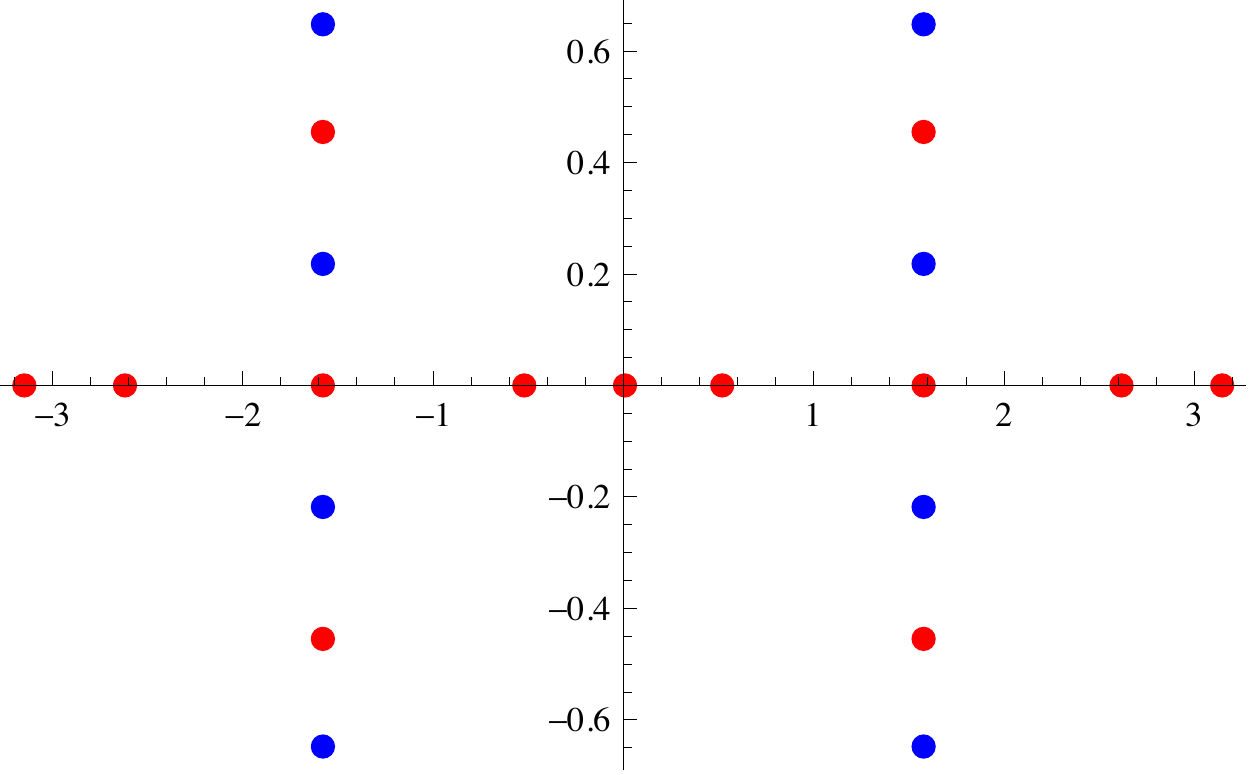}
\caption{Vortex equilibria in the Whittaker--Hill case with $s=5$, $\alpha=1.5$ and $J=\{4,5\}$, $I=\{4\}$ (left),
$J=\{1,5\}$, $I=\{1\}$ (right).}
\label{WH3}
\end{figure}

Two examples of vortex equilibria in the case $s=5$ are shown in Fig.~\ref{WH3}.

\section{Doubly-periodic vortex equilibria}

The classical Lam\'{e} operator has the form
\begin{gather*}
L = -D^2 + s(s+1) \wp(z),
%\label{Lame}
\end{gather*}
where $\wp(z)$ is the Weierstrass elliptic function and~$s$ is an integer.
For the Lam\'{e} operator there are explicit formulae for the $2s+1$ eigenfunctions, known as Lam\'e functions, going
back to Hermite~\cite{WW}.
Replacing in the previous section $\psi_i(z)$ by the Lam\'e functions we have new doubly periodic vortex equilibria.

In fact, we have the following very general result.
Let $\sigma(z)$, $\zeta(z)$, $\wp(z)$ be classical Weierstrass functions with periods $2\omega_1$, $2\omega_2$, see
Whittaker--Watson~\cite{WW}.

Suppose that
\begin{gather*}
%\label{psi1}
\psi(z)=\prod\limits_{i=1}^{N} \sigma^{m_i}(z-z_i) e^{Bz}
\end{gather*}
with integer $m_i$ with zero sum, is a~solution of the Schr\"odinger equation
\begin{gather*}
%\label{eq1}
-\psi''+u(z)\psi=E\psi
\end{gather*}
with the potential
\begin{gather*}
u(z)=\sum\limits_{i=1}^N m_i(m_i-1)\wp(z-z_i).
\end{gather*}
Then we claim that $W(z)=\frac{1}{2\pi i}\log \psi(z)$ is the complex potential of a~relative doubly periodic vortex
equilibrium in the moving frame.
Indeed, the function
\begin{gather*}
f=D \log\psi=\sum\limits_{i=1}^N m_i\zeta(z-z_i)+B
\end{gather*}
must satisfy the Riccati equation
\begin{gather*}
f'+f^2=u(z)-E.
\end{gather*}
Since the residues of the potential are zero, the same must be true for $f^2$, which implies the Stieltjes conditions
\begin{gather}
\label{Stil}
\sum\limits_{j\neq i}^Nm_j \zeta(z_i-z_j) +B=0,
\qquad
i=1,\dots, N.
\end{gather}
Now recall the equations for the dynamics of the doubly periodic conf\/iguration of vortices.
Let $z_1, \dots, z_N$ with circulations $\Gamma_1, \dots, \Gamma_N$ be part of such conf\/iguration within the
parallelogram def\/ined by the periods $2\omega_1$, $2\omega_2$. Then their dynamics is described by the equations
\begin{gather*}
%\label{eqper}
\frac{d\bar z_k}{dt}=\frac{1}{2\pi i}\left(\sum\limits_{j\neq k}^N\Gamma_j \zeta(z_k-z_j) +C\right),
\qquad
i=1,\dots, N,
\end{gather*}
where
\begin{gather*}
%\label{C}
C=a\sum\limits_{j=1}^N\Gamma_j z_j+b\sum\limits_{j=1}^N\Gamma_j \bar z_j
\end{gather*}
and~$a$,~$b$ are solutions of the linear system
\begin{gather*}
a \omega_i+b \bar \omega_i=\eta_i,
\qquad
i=1,2,
\end{gather*}
with the usual notation $\eta_i=\zeta(\omega_i)$ (see~\cite{AS2, Tkac}).
If $z_1, \dots, z_N$ are the zeros of~$\psi$ and $\Gamma_i=m_i$, then due to Stieltjes relations~\eqref{Stil} we have
\begin{gather*}
\frac{d\bar z_k}{dt}=\frac{1}{2\pi i}(C-B)=A,
\end{gather*}
which shows that the conf\/iguration of zeros of~$\psi$ is indeed a~relative vortex equilibrium moving with velocity
$v=\bar A$.
The complex potential of the f\/low in the moving frame is $W=\frac{1}{2\pi i} \log \psi(z)$, the instantaneous one in the
f\/ixed frame is
\begin{gather*}
W_*=W+Az=\frac{1}{2\pi i} \log\big(\psi e^{(C-B)z}\big).
\end{gather*}

Conversely, assume that we have a~doubly periodic relative vortex equilibrium $z_1, \dots, z_N$ with integer
circulations $\Gamma_i=m_i$, so that
\begin{gather*}
\frac{d\bar z_k}{dt}=\frac{1}{2\pi i}\left(\sum\limits_{j\neq k}^N\Gamma_j \zeta(z_k-z_j) +C\right)=A,
\end{gather*}
then we have Stieltjes relations~\eqref{Stil} with $B=C-2\pi i A$, which guarantee the Riccati relation (since the
functions are elliptic) and hence the Schr\"odinger equation of the required form.
This proves Theorem~\ref{Theorem2}.

Of course, in that form the claim becomes almost a~tautology since both conditions are equivalent to the Stieltjes
relations~\eqref{Stil}, which it is not clear how to solve.
Fortunately we have several concrete examples coming from the theory of elliptic solitons~\cite{AMM, DN, GW, Krichever,
S, T, Take, TV}, which will be discussed elsewhere.

Here we only mention classical Hermite's solution of the Lam\'e equation $L\psi=E \psi$
\begin{gather*}
%\label{Herm}
\psi=\prod\limits_{r=1}^s\frac{\sigma(a_r-z)}{\sigma(z)\sigma(a_r)}e^{B z},
\qquad
B=\sum\limits_{r=1}^s \zeta(a_r),
\end{gather*}
which gives doubly periodic vortex streets with $z=0$ of circulation $-s$ and $z_r=a_r$, $r=1,\dots, s$ of circulation~1,
depending on an arbitrary parameter~$E$ (see the details in Whittaker and Watson~\cite[Section~23.7]{WW}).

\section{Examples: pictures and analysis}\label{Section6}

We restrict ourselves with the analysis of the vortex conf\/igurations given by Theorem~\ref{Theorem1}.
As we will see in spite of the simple explicit formulae there are many natural questions to answer already in this case.

As we have seen above, in the simplest case $n=1$ we have the original von K\'arm\'an vortex streets~\cite{Lamb, Kar,Kar2}.

\subsection[Case $n=2$ with $k_1=1$, $k_2=2$]{Case $\boldsymbol{n=2}$ with $\boldsymbol{k_1=1}$, $\boldsymbol{k_2=2}$}

\begin{figure}[h] \centering
\includegraphics[width=7.5cm]{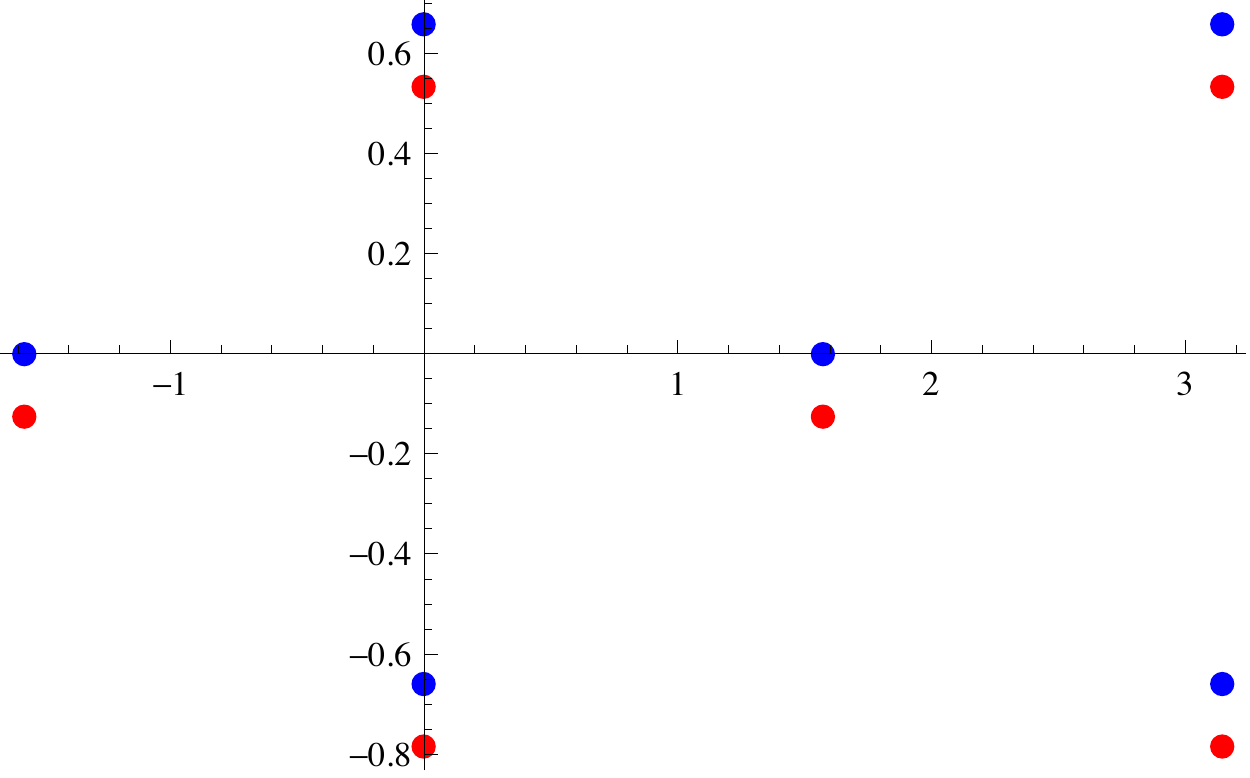}
 \caption{Conf\/iguration $\Sigma_{k,\phi,\kappa}$ with $k=(1,2)$,
$\phi =(0,\pi/2)$ and $\kappa=8$.
The circulations of the red and blue vortices are $\Gamma = 1$ and $\Gamma=-1$ respectively.}
\label{3Karman}
\end{figure}

An example of the corresponding vortex street is shown at Fig.~\ref{3Karman}.

Let us assume that $\phi_1=\phi_2=0$. As we will see in that case because of the cancelation we have only three
vortices per period: one with circulation~$-2$ (located at~0) and two with circulation~$+1$.

Indeed in this case $W(\sin z, \sin 2z) = -2\sin^3 z$ and
\begin{gather*}
W\big(\sin z, \sin 2z,e^{i\kappa z}\big)=-\sin z e^{i \kappa z} \big(\big(2+\kappa^2\big)\cos 2z - 3i\kappa \sin 2z + \big(4-\kappa^2\big)\big),
\end{gather*}
so we have the cancelation of $\sin z$ and
\begin{gather*}
%\label{cp}
\psi = \frac{W(\sin z, \sin 2z,e^{i\kappa z})}{W(\sin z, \sin 2z)}
=\frac{\big(2+\kappa^2\big)\cos 2z - 3i\kappa \sin 2z+\big(4-\kappa^2\big)}{2 \sin^2 z} e^{i \kappa z}.
\end{gather*}
The equation
\begin{gather*}
\big(2+\kappa^2\big)\cos 2z - 3i\kappa \sin 2z + \big(4-\kappa^2\big) = 0
\end{gather*}
can be rewritten as
\begin{gather*}
(\kappa-1)(\kappa-2)e^{2iz} + (\kappa+1)(\kappa+2)e^{-2iz}-2(\kappa+2)(\kappa-2)=0,
\end{gather*}
which is a~quadratic in $X = e^{2iz}$ with roots
\begin{gather*}
X_{1,2}= \frac{\big(\kappa^2-4\big) \pm \sqrt{3\big(4-\kappa^2\big)}}{(\kappa-1)(\kappa-2)}.
\end{gather*}
So, we have a~relative equilibrium with the vortices at the following locations (and~$\pi$-periodically)
\begin{gather*}
z_0=0
\qquad
\textrm{with}
\qquad
\Gamma = -2,
\\
z_{1,2}=\frac{1}{2i} \log \left(\frac{(\kappa^2-4) \pm \sqrt{3(4-\kappa^2)}}{(\kappa-1)(\kappa-2)}\right)
\qquad
\textrm{with}
\qquad
\Gamma = 1
\end{gather*}
moving with the velocity $v=-\bar\kappa/2\pi$ for non-critical $\kappa \neq 1,2$.

\begin{figure}[h] \centering
\includegraphics[width=6cm]{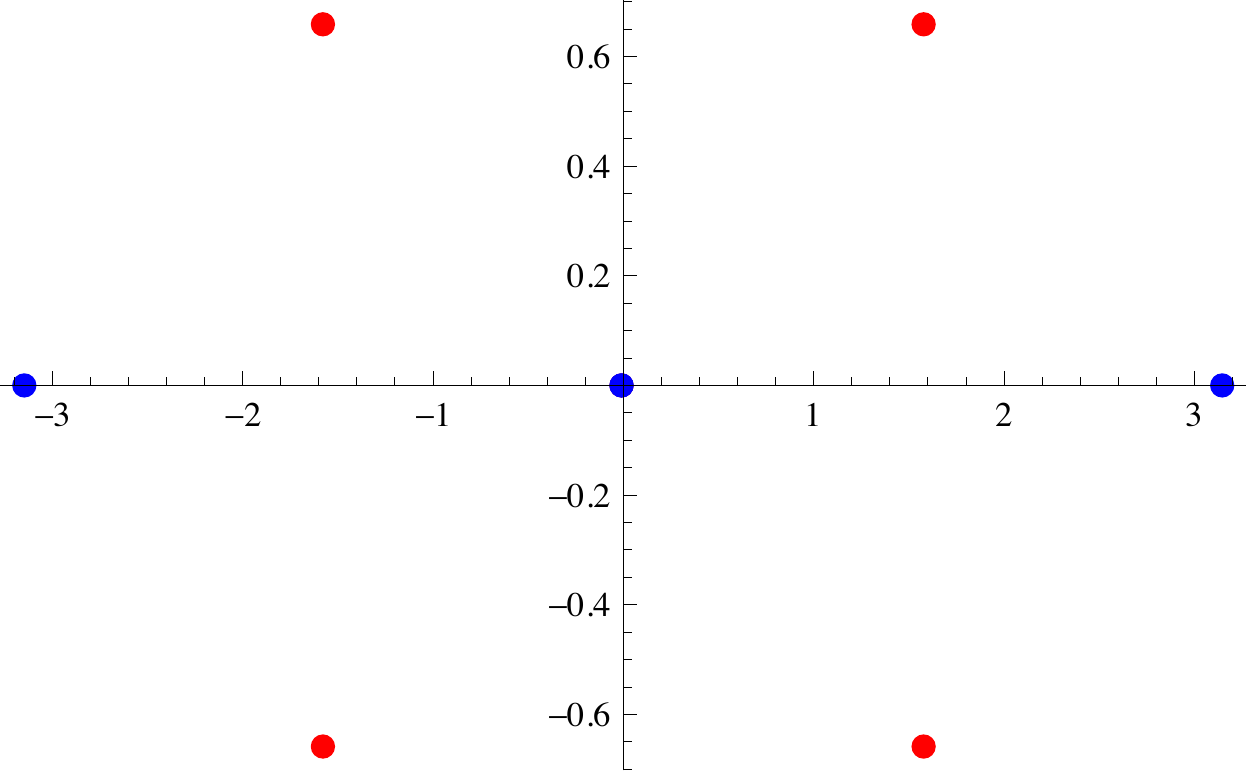} \hspace{1pt} \includegraphics[width=6cm]{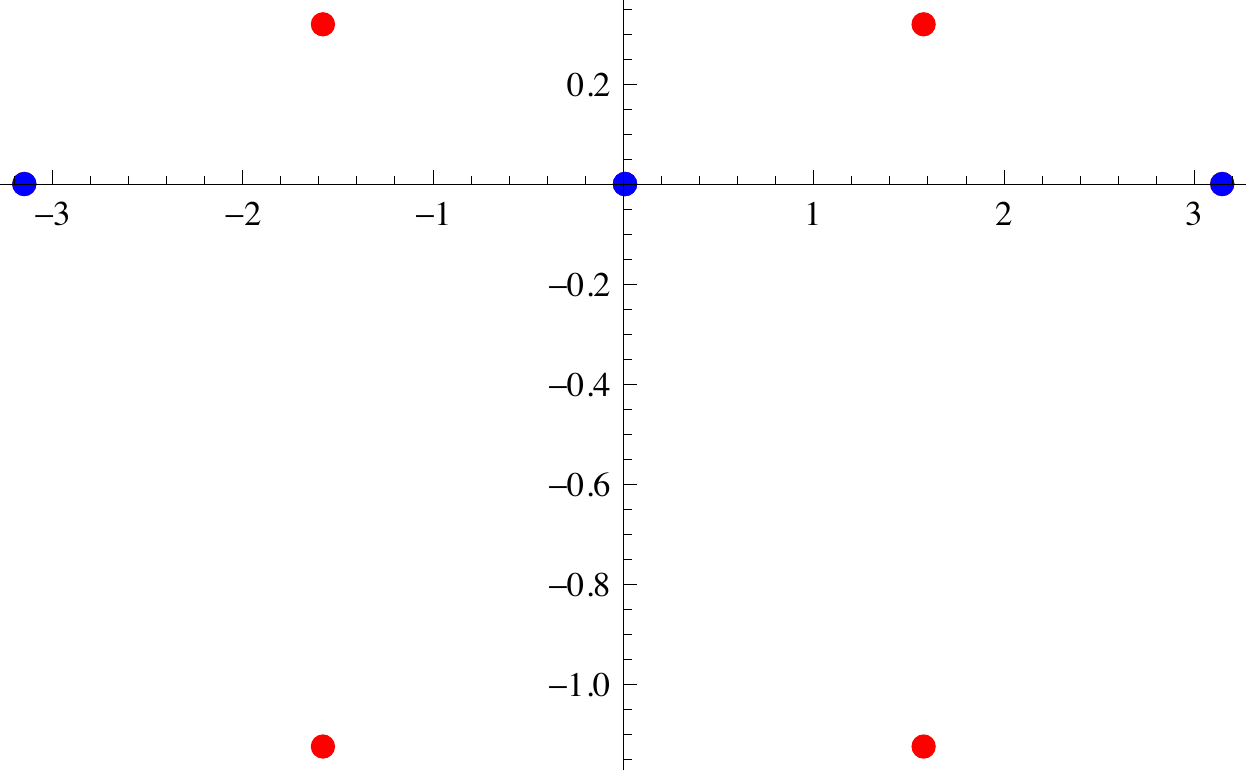}

\caption{$\Sigma_{k,0,\kappa}$ with $k=(1,2)$ and $\kappa=0$ (equilibrium, L), $\kappa=0.5$ (R).
Circulation of the f\/ixed vortex at $0$ is $-2$, while the other vortices have circulation 1.}
\label{WKCritical1}
\end{figure}

\begin{figure}[h] \centering \includegraphics[width=6cm]{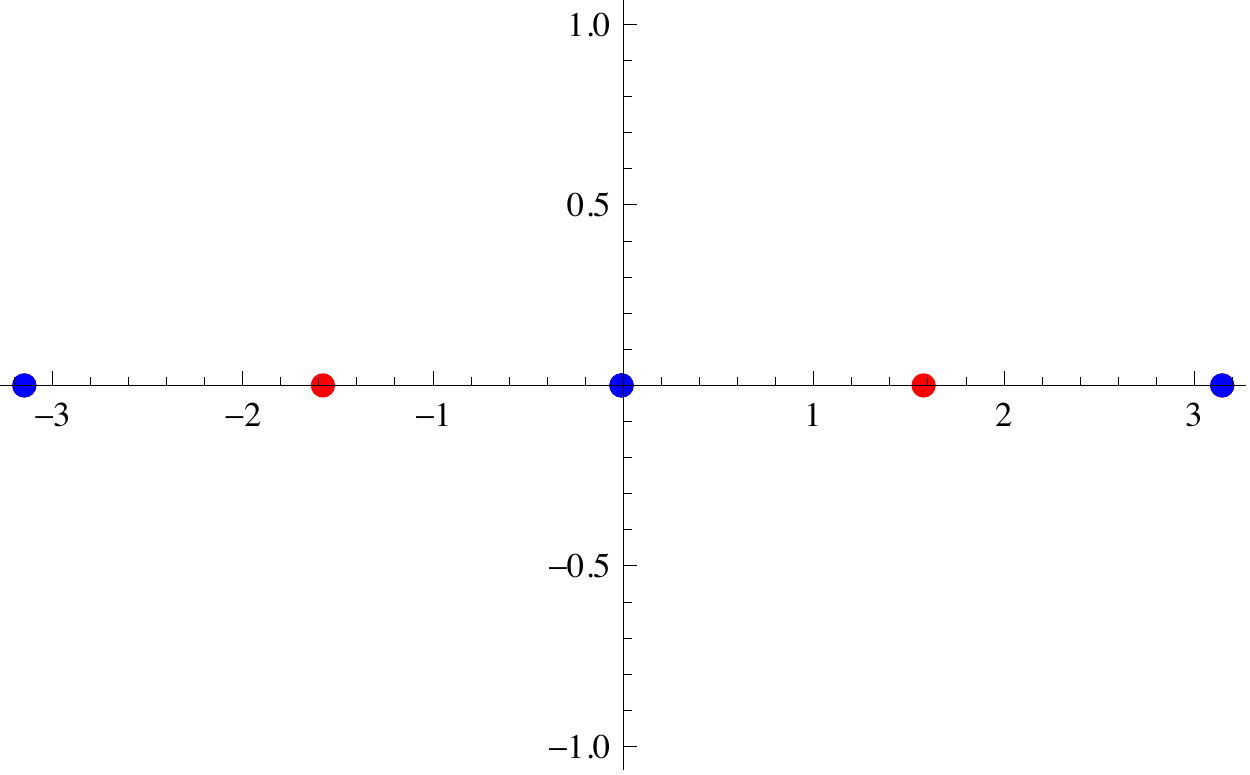} \hspace{1pt} \includegraphics[width=6cm]{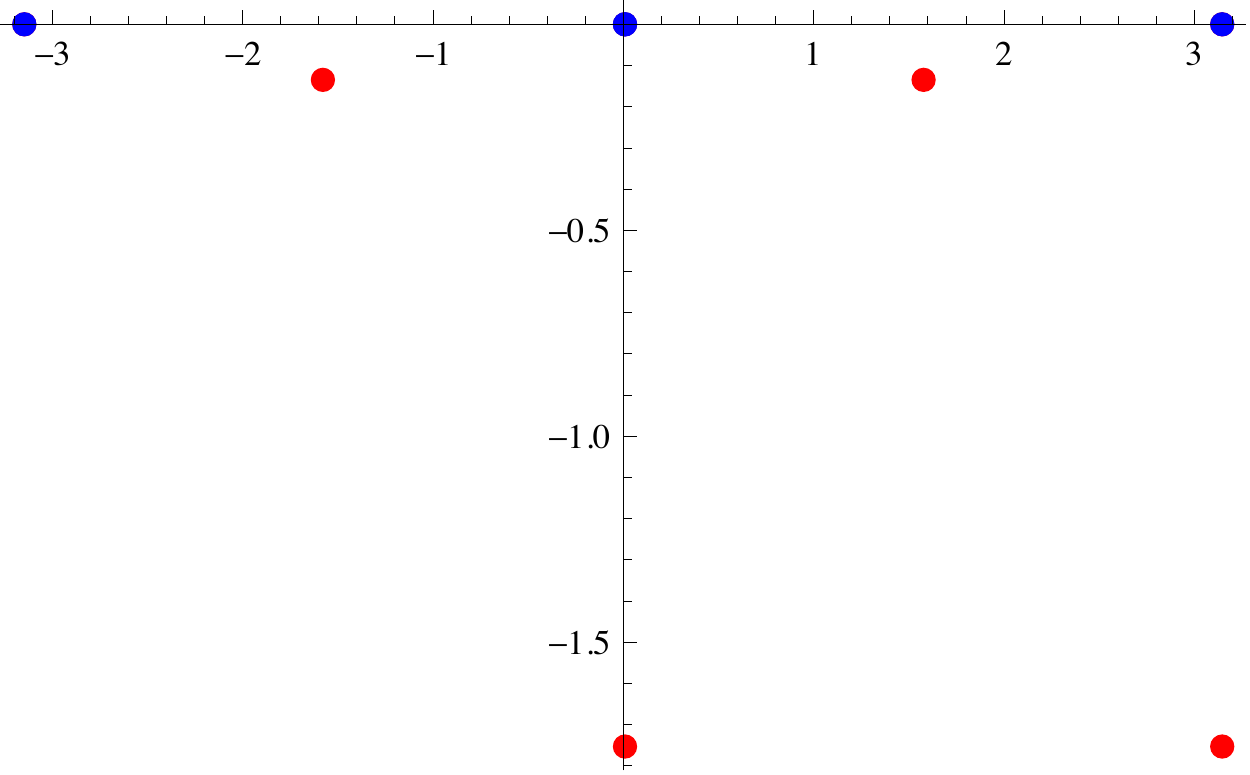}
\caption{$\Sigma_{k,0,\kappa}$ with $k=(1,2)$ and $\kappa=1$ (critical, L), $\kappa=1.2$ (R).}
\label{WKCritical2}
\end{figure}

\begin{figure}[h] \centering \includegraphics[width=6cm]{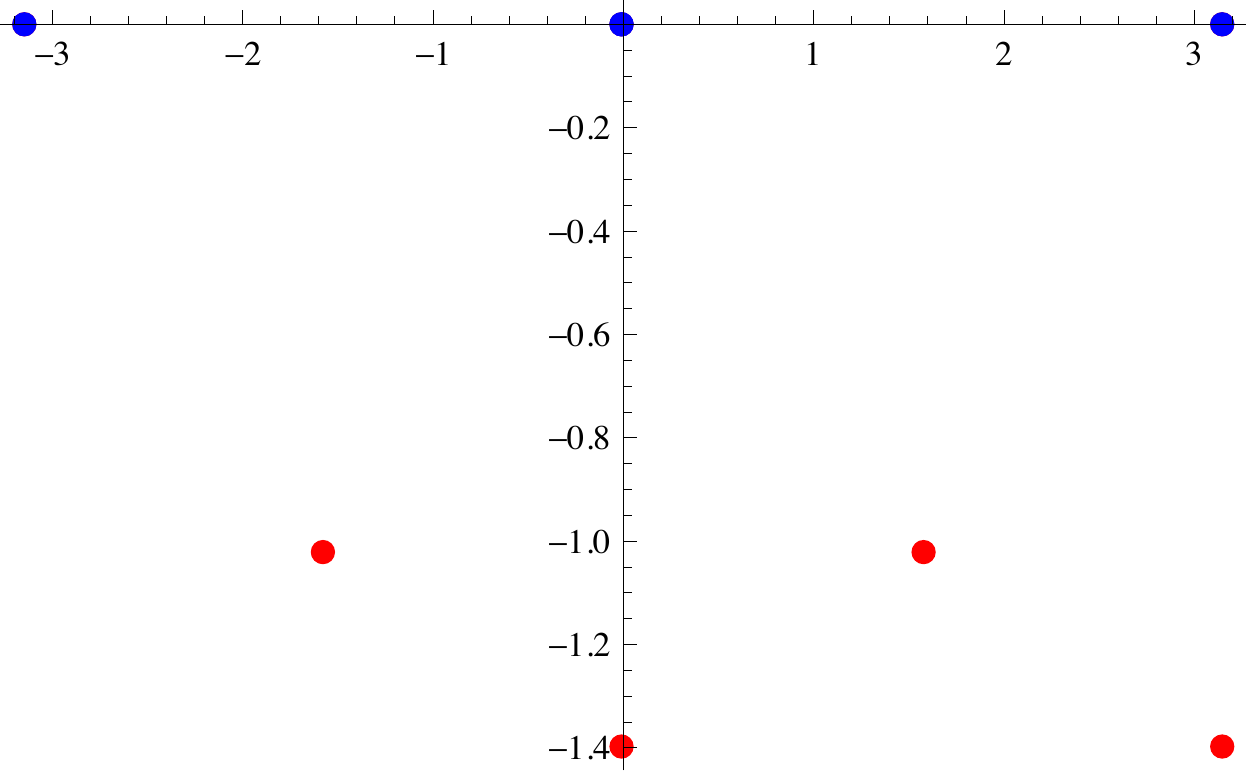} \hspace{1pt} \includegraphics[width=6cm]{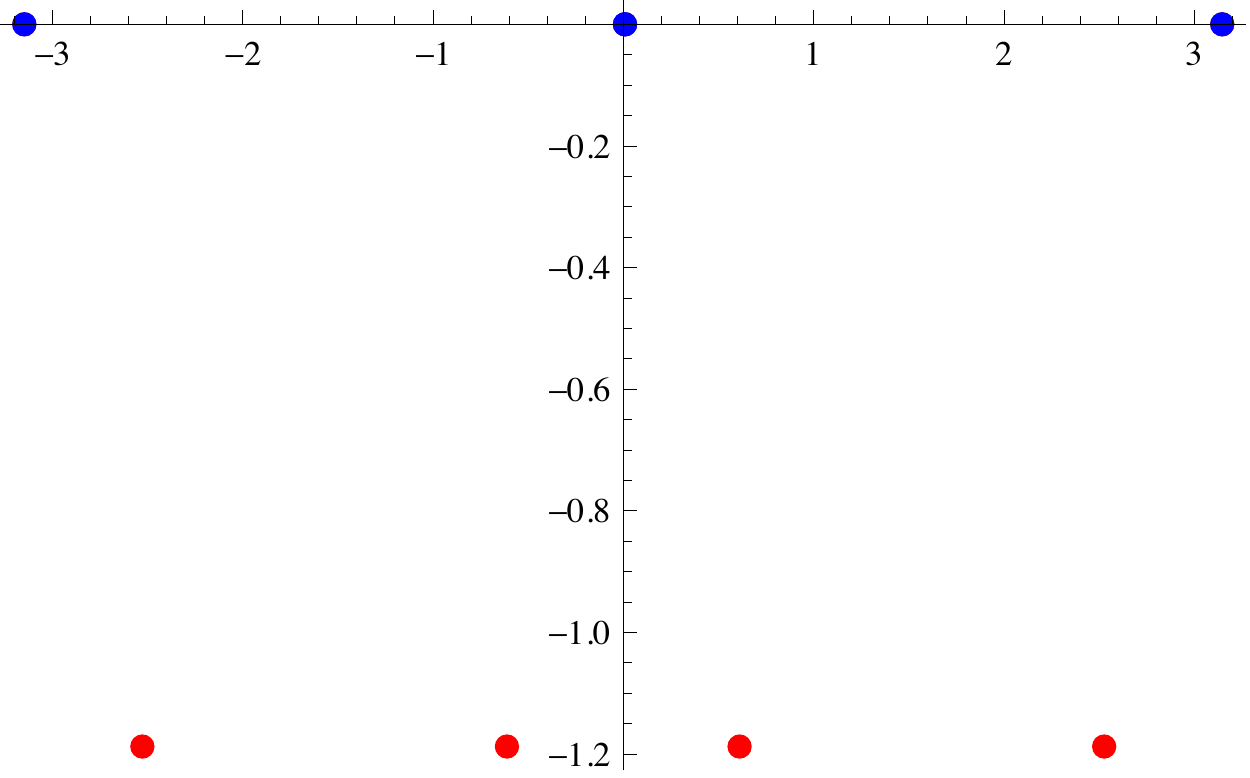}

\caption{$\Sigma_{k,0,\kappa}$ with $k=(1,2)$ and $\kappa=1.9$ (L), $\kappa=2.1$ (R).}
\label{WKCritical3}
\end{figure}

\begin{figure}[h] \centering \includegraphics[width=6cm]{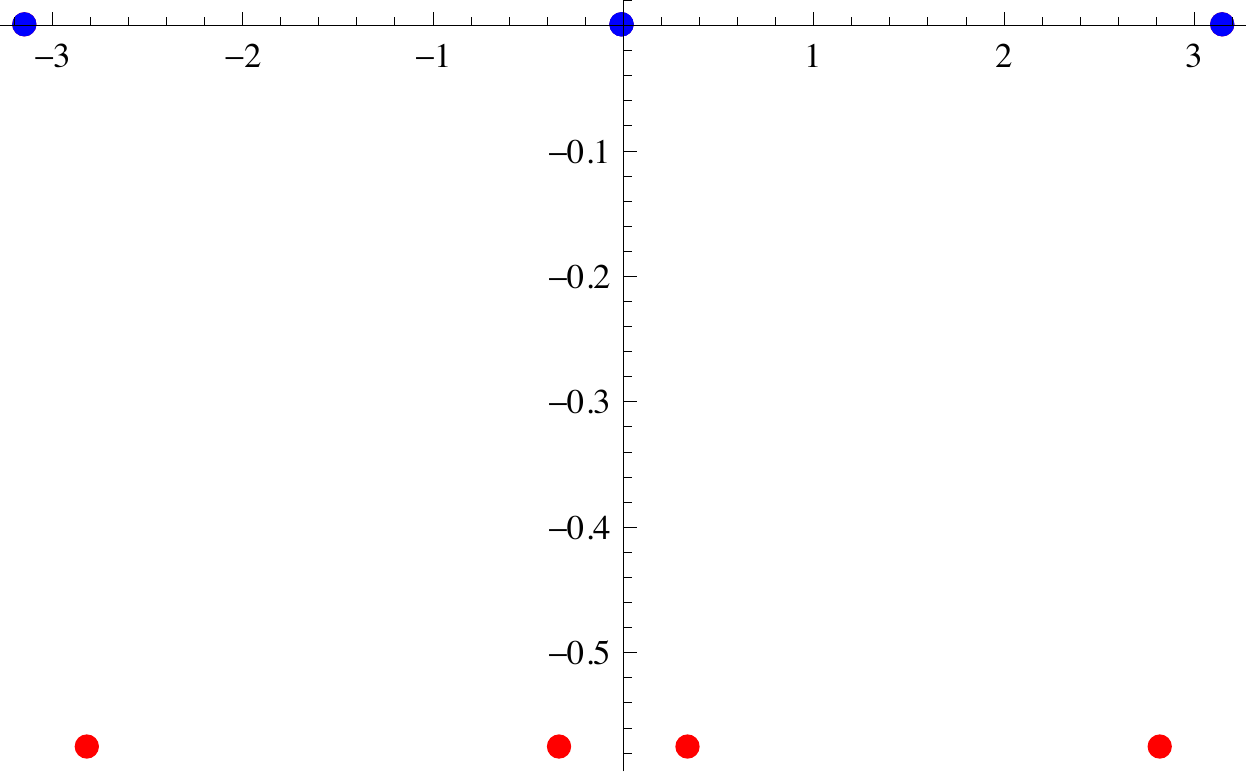} \hspace{1pt} \includegraphics[width=6cm]{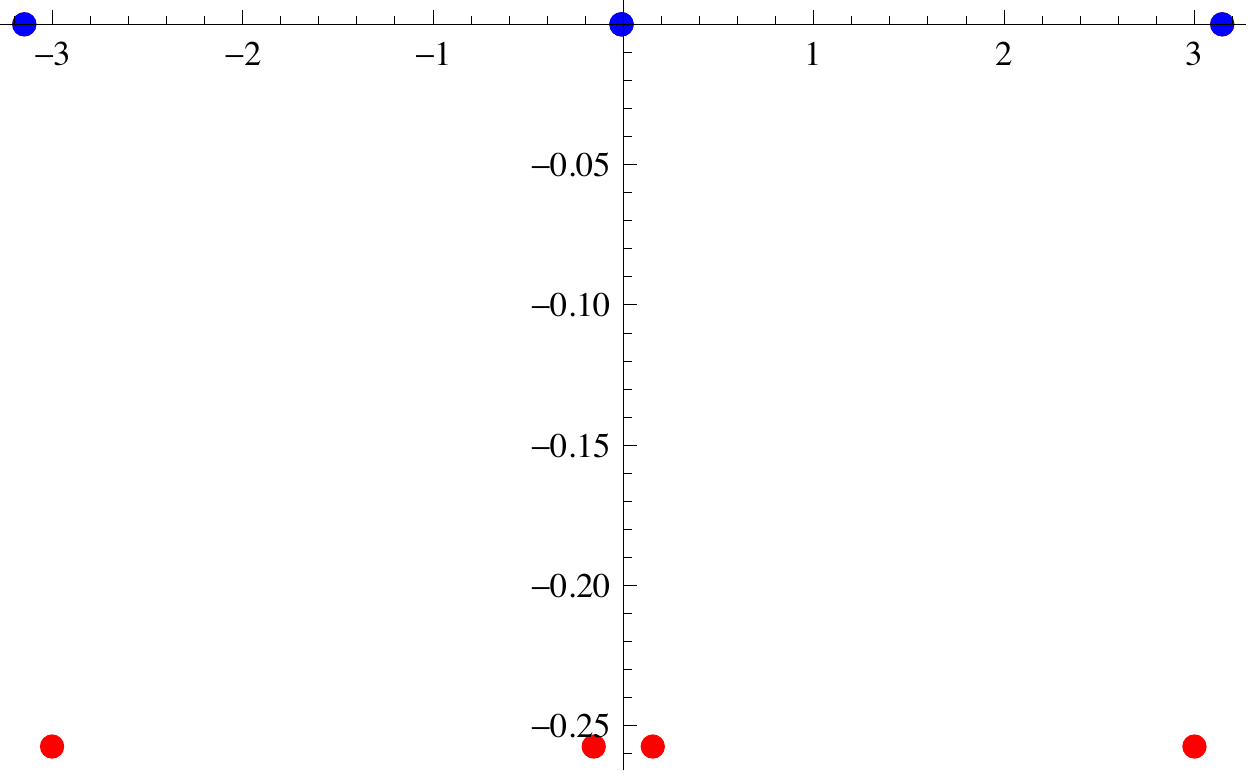}

\caption{$\Sigma_{k,0,\kappa}$ with $k=(1,2)$ and $\kappa=3$ (L), $\kappa=6$ (R)}
\label{WKCritical4}
\end{figure}

When $\kappa=0$ we have $X_{1,2} = -2 \pm \sqrt{3}$ and an equilibrium conf\/iguration shown on the very left of
Fig.~\ref{WKCritical1} with red vortices $z_{1,2} = \pi/2 \pm \frac{i}{2}\log (2+\sqrt{3})$.
Figs.~\ref{WKCritical1}--\ref{WKCritical4} show what happens when the parameter~$\kappa$ grows.

Let us look at what happens near the critical values $\kappa=1$ and $\kappa =2$.
Let $\kappa = 1+ \epsilon$ with small $\epsilon$, then
\begin{gather*}
z_{1,2} \approx \frac{1}{2i} \log \left(\frac{(-3+2\epsilon) \pm (3-\epsilon)}{-\epsilon}\right),
\end{gather*}
so the limiting value is $z_1 = \frac{1}{2i} \log (-1) = \frac{1}{2i} \log e^{i\pi} =\frac{\pi}{2}$ while $z_2 \approx
\frac{1}{2i} \log (6/\epsilon) \to -i\infty$ goes to inf\/inity.
At the critical value $\kappa =1$ we have a~genuine equilibrium shown
at Fig.~\ref{WKCritical2} with~$0$ and~$\pi/2$ having circulations~$-2$ and~$1$ respectively
corresponding to $\psi=\sin 2z/\sin^3z=2\cos z/\sin^2z$.

Similarly, setting $\kappa = 2+ \epsilon$, $\epsilon >0$ small
\begin{gather*}
z_{1,2} \approx \frac{1}{2i} \log \left(\frac{4\epsilon \pm (-12\epsilon)^{1/2}}{\epsilon}\right)
\approx \frac{1}{2i} \log\big(\pm 2i(3/\epsilon)^{1/2}\big),
\end{gather*}
so in the limit $\epsilon \to 0$ $z_{1,2}$ go to $-i\infty$ with the real parts approaching $\pm \pi/4$.
If instead we let $\kappa = 2- \epsilon$, $\epsilon >0$, then
\begin{gather*}
z_{1,2} \approx \frac{1}{2i} \log \left(\frac{-4\epsilon \pm (12\epsilon)^{1/2}}{-\epsilon}\right)
= \frac{1}{2i} \log\big(4 \mp 2(3/\epsilon)^{1/2}\big),
\end{gather*}
so this time in the limit $\epsilon \to 0$ the real parts of $z_{1,2}$ approach $0$ and $\pi/2$ (see
Fig.~\ref{WKCritical3}).
At the critical level $\kappa=2$ we have the trivial equilibrium with vortices of circulation~$-2$ at~0 corresponding to
$\psi=\sin z/\sin^3z=\sin^{-2}z$ with both vortices with circulation~1 gone to inf\/inity.

When $\kappa \to\infty$, then $X_{1,2}\to 1$ so that red vortices $z_{1,2} \to n\pi$, $n\in\mathbb Z$ approach blue one
(see Fig.~\ref{WKCritical4}).
For $\kappa<0$ the conf\/iguration will look the same as the corresponding positive conf\/iguration but will move to the
left, instead.

\subsection[Case $n=2$ with $k_1=m$, $k_2=n$, $\phi_1=\phi_2=0$]{Case $\boldsymbol{n=2}$ with $\boldsymbol{k_1=m}$,
$\boldsymbol{k_2=n}$, $\boldsymbol{\phi_1=\phi_2=0}$}

The corresponding conf\/iguration of vortices $\Sigma_{m,n,\kappa}$ is given by the zeros of
\begin{gather}
W_{m,n} = W(\sin mz, \sin nz) = ((n-m)\sin (m+n)z - (m+n)\sin (n-m)z)/2,
\nonumber
\\
W_{m,n}^\kappa = W(\sin mz,\sin nz, e^{i\kappa z})
\nonumber
\\
\phantom{W_{m,n}^\kappa}{}
=\big((n+m)\big(\kappa^2-mn\big) \sin (n-m)z-(n-m)\big(\kappa^2+mn\big) \sin (n+m)z
\nonumber
\\
\phantom{W_{m,n}^\kappa =}{}
+i\kappa\big(n^2-m^2\big)(\cos (n-m)z - \cos(n+m)z)\big)e^{i\kappa z}/2.
\label{Eqq}
\end{gather}

When $\kappa=0$ we have a~genuine equilibrium.
In that case $W_{m,n}^0 = mn W(\cos mz, \cos nz)$, where
\begin{gather*}
W(\cos mz, \cos nz) = \frac{n-m}{2} \sin (n+m)z + \frac{n+m}{2} \sin (n-m)z.
\end{gather*}
The number of common zeros of $W_{m,n}^0$ and $W_{m,n}$ depends on arithmetic of~$m$ and~$n$.
If~$d$ is the greatest common divisor of~$m$ and~$n$ then $W_{m,n}^0$ and $W_{m,n}$ have~$d$ common zeros at $z=l\pi/d$,
 $l=0,\dots,d-1$ of multiplicities~1 and~3 respectively.
Thus we have per period $m+n-3d$ vortices of circulation~$-1$,
$d$ vortices of circulation~$-2$ and $m+n-d$ vortices of
circulation~1 (see the example with $m=10$, $n=12$ below).
The same is true for generic values of~$\kappa$.

\begin{figure}[h] \centering \includegraphics[width=10cm]{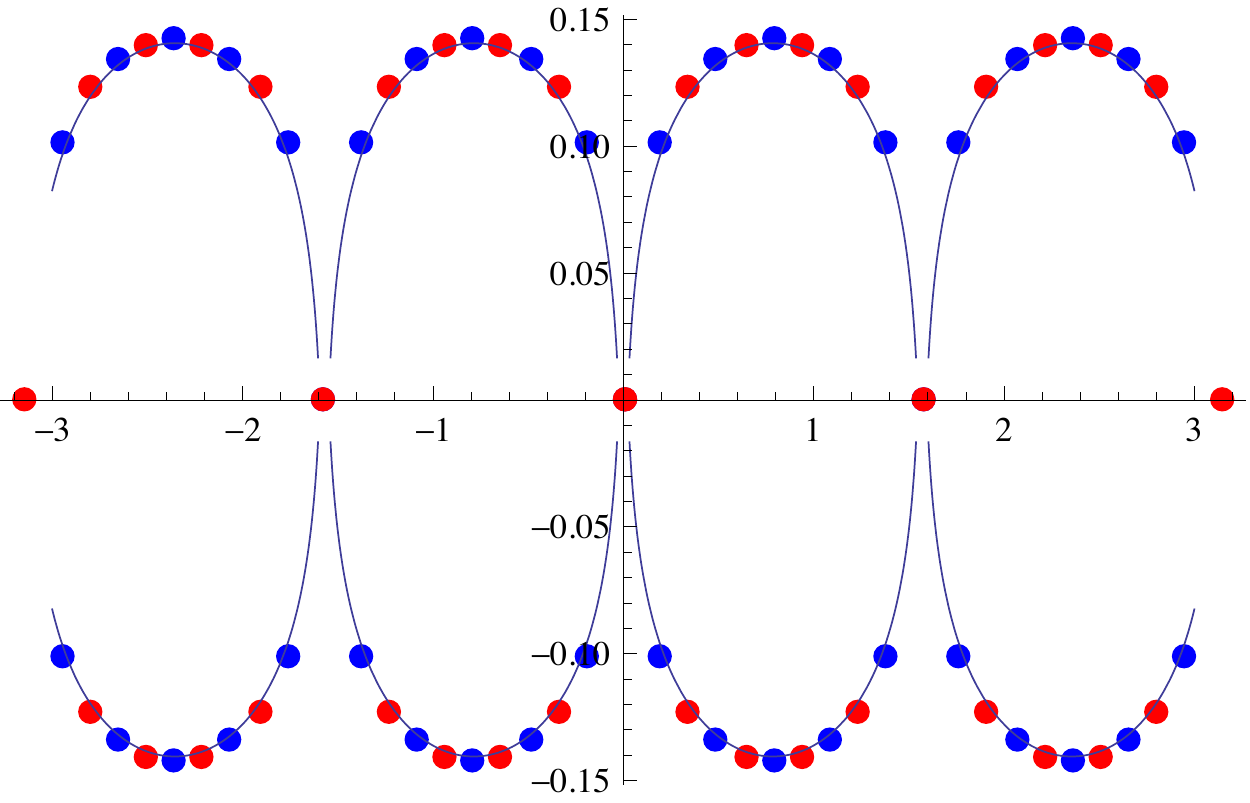}

\caption{Vortex equilibrium conf\/iguration $\Sigma_{10,12,0,0}$
against the asymptotic curve~\eqref{triglimcurve}.}
\label{WK5}
\end{figure}

Since the picture suggests that the vortices lie on some curve let us try to f\/ind its shape.
Our arguments here are similar to the analysis of the Wronskians of Hermite polynomials in~\cite{FHV}.

For complex zeros of both $W_{m,n}^0$ and $W_{m,n}$ we have
\begin{gather}
\label{trigcurve}
|\sin (m+n)z| = \frac{m+n}{n-m} |\sin (n-m)z|.
\end{gather}
Let us assume that~$m$,~$n$ are large compared with the dif\/ference $m-n$. Then in the upper half-plane $z=x+iy$, $y>0$
the negative exponential term in $\sin (m+n)z$, with modulus $e^{(m+n)y}/2$, will dominate.
Taking the modulus of both sides and assuming that~$y$ is small, so that $\sin(n-m)z \approx \sin(n-m)x$,
equation~\eqref{trigcurve} becomes
\begin{gather*}
\frac{1}{2}e^{(m+n)y} \approx \frac{m+n}{n-m} |\sin (n-m)x|.
\end{gather*}
Taking logarithms and combining with the lower half-plane case, we have the following appro\-xi\-ma\-te formula for the curve
on which the zeros lie
\begin{gather}
\label{triglimcurve}
|y| = \pm \frac{1}{m+n} \left(\log|\sin (n-m)x| +\log \frac{2(m+n)}{n-m}\right).
\end{gather}
Fig.~\ref{WK5} shows a~good agreement with this formula already for $m=7$, $n=8$.

When the parameter~$\kappa$ (which is essentially velocity) increases from zero the red vortices lie on their own
independent curves.
We will now derive a~formula for these curves, from equation~\eqref{Eqq}.
Setting $W_{m,n}^\kappa=0$, then collecting terms with argument $(n+m)z$ onto the left and those with argument $(n-m)z$
onto the right, we have
\begin{gather*}
(n-m)\big(\kappa(n+m) \cos(n+m)z -i\big(\kappa^2+mn\big) \sin(n+m)z\big)
\\
\qquad
=(n+m)\big(\kappa(n-m)\cos (n-m)z-i\big(\kappa^2-mn\big) \sin(n-m)z\big).
\end{gather*}
Writing the left hand side in terms of exponentials, we get
\begin{gather*}
\frac{n-m}{2}\big((\kappa-n)(\kappa-m)e^{i(n+m)z} - (\kappa+n)(\kappa+m)e^{-i (n+m)z}\big)
\\
\qquad
=(n+m)\big(\kappa(n-m)\cos (n-m)z - i \big(\kappa^2-mn\big) \sin (n-m)z\big).
\end{gather*}
Since $(n+m)$ is assumed to be large, in the upper half-plane the negative exponential term will dominate.
Set $z=x+iy$ and assume that~$y$ is small enough that $\sin z \approx \sin x$ and similarly for $\cos z$.
Taking the modulus of both sides we have
\begin{gather*}
\frac{n-m}{2}(\kappa+n)(\kappa+m)e^{(n+m)y}
\\
\qquad
\approx(n+m)\big(\big(\kappa^2-n^2\big)\big(\kappa^2-m^2\big)\sin^2(m-n)x +\kappa^2(n-m)^2\big)^{1/2}.
\end{gather*}
Taking logarithms of both sides we arrive at the formula
\begin{gather*}
y = \frac{1}{2(n+m)} \log\left|\frac{\kappa-n}{\kappa+n}\frac{\kappa-m}{\kappa+m}\right| + \frac{1}{(n+m)} \log \frac{2(n+m)}{n-m}
\\
\phantom{y=}
+ \frac{1}{2(n+m)}\log\left|\sin^2 (n-m)x + \frac{\kappa^2(n-m)^2}{(\kappa^2-n^2)(\kappa^2-m^2)}\right|,
\end{gather*}
and by a~similar calculation in the lower half-plane (where we keep the positive exponential term instead) we have
a~formula for the lower line of vortices.
Combining the two gives us
\begin{gather}
y_{\pm} = \frac{1}{2(n+m)} \log\left|\frac{\kappa-n}{\kappa+n}\frac{\kappa-m}{\kappa+m}\right|
\nonumber
\\
\phantom{y_{\pm}=}
\pm \frac{1}{2(n+m)}\left(2\log\frac{2(n+m)}{n-m}+\log\left|\sin^2(n-m)x + \frac{\kappa^2(n-m)^2}{(\kappa^2-n^2)(\kappa^2-m^2)}\right|\right).
\label{movingcurve}
\end{gather}

\begin{figure}[h!] \centering \includegraphics[width=5cm]{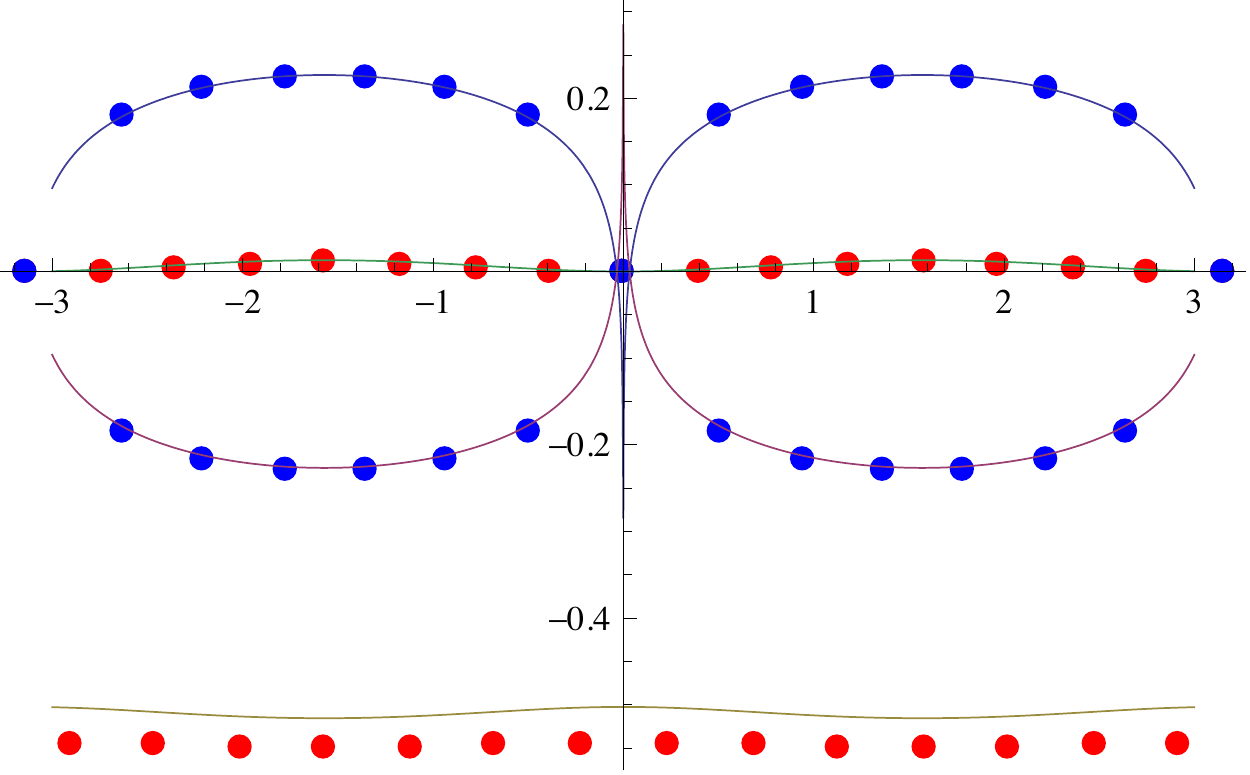} \hspace{1pt} \includegraphics[width=5cm]{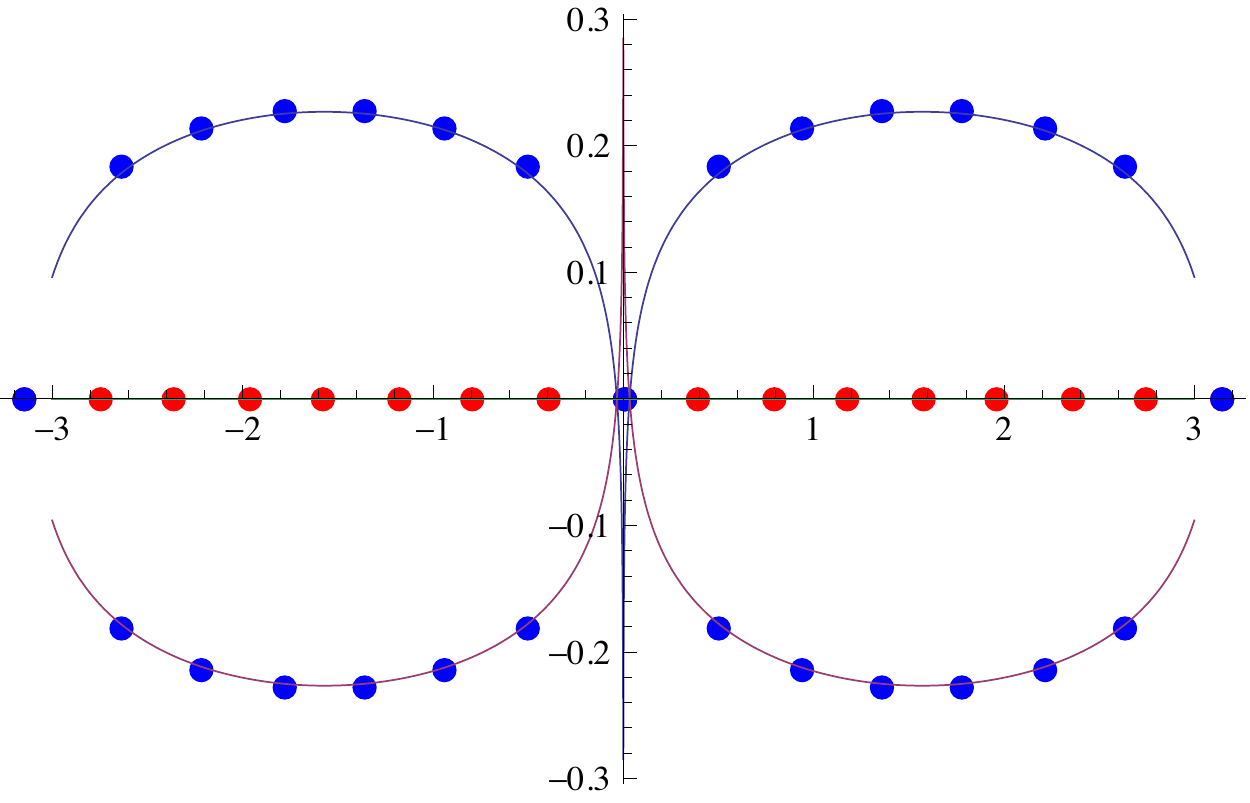} \hspace{1pt}
\includegraphics[width=5cm]{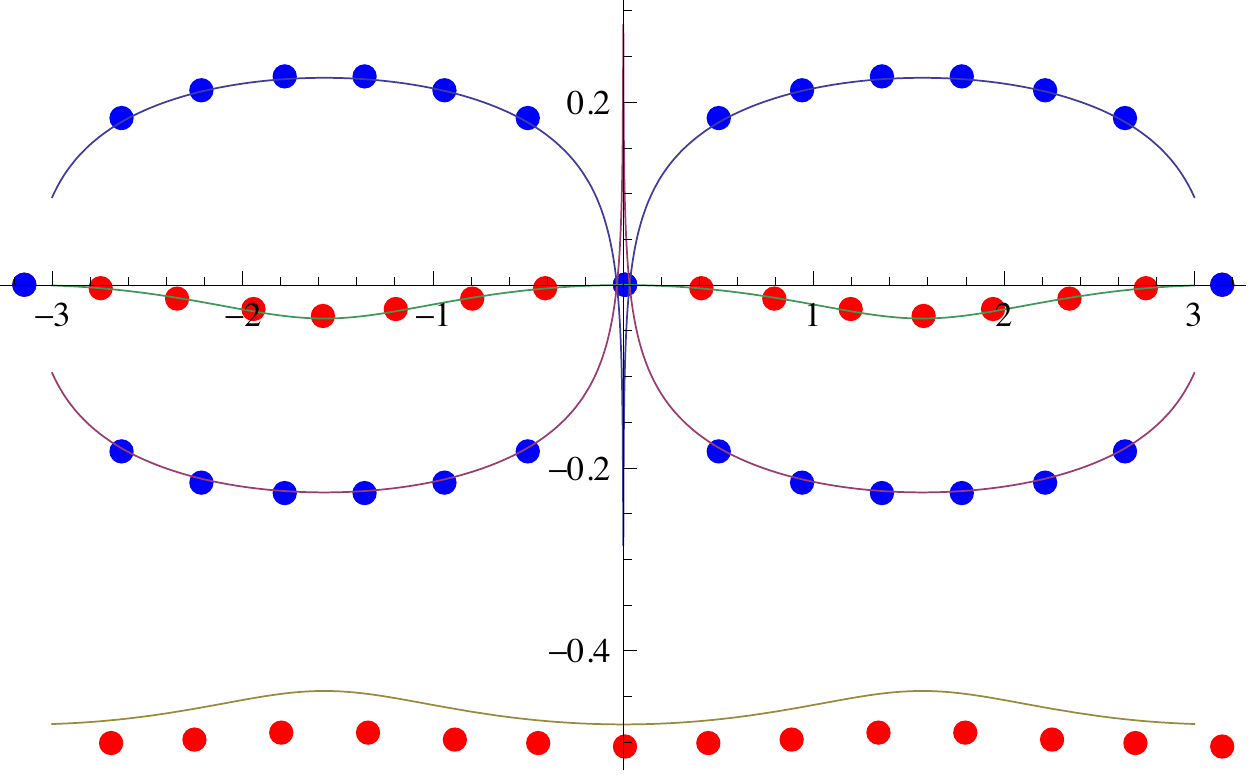}

\caption{$\Sigma_{7,8,\kappa}$ with $\kappa=6.9$ (L), $\kappa=7$ (critical, M),
$\kappa=7.2$ (R).}
\label{WK7}
\end{figure}

\begin{figure}[h!] \centering \includegraphics[width=5cm]{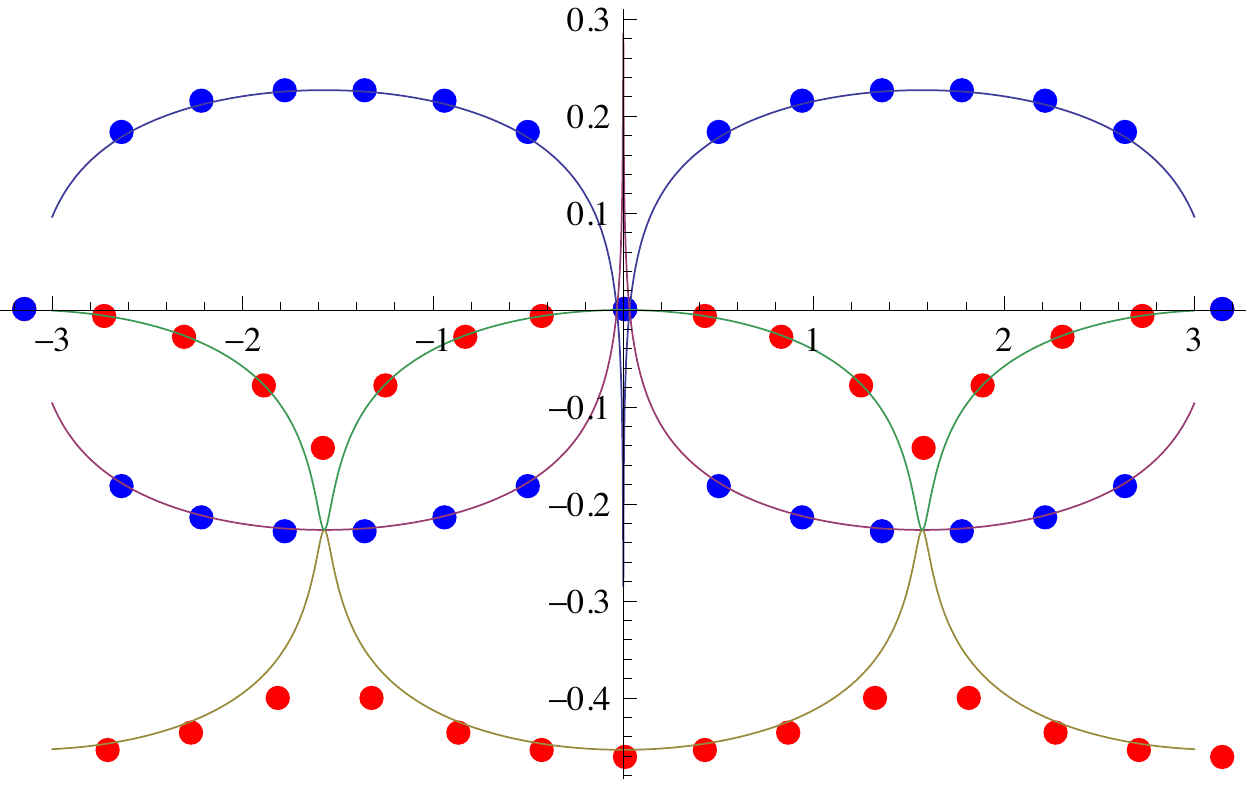} \hspace{1pt} \includegraphics[width=5cm]{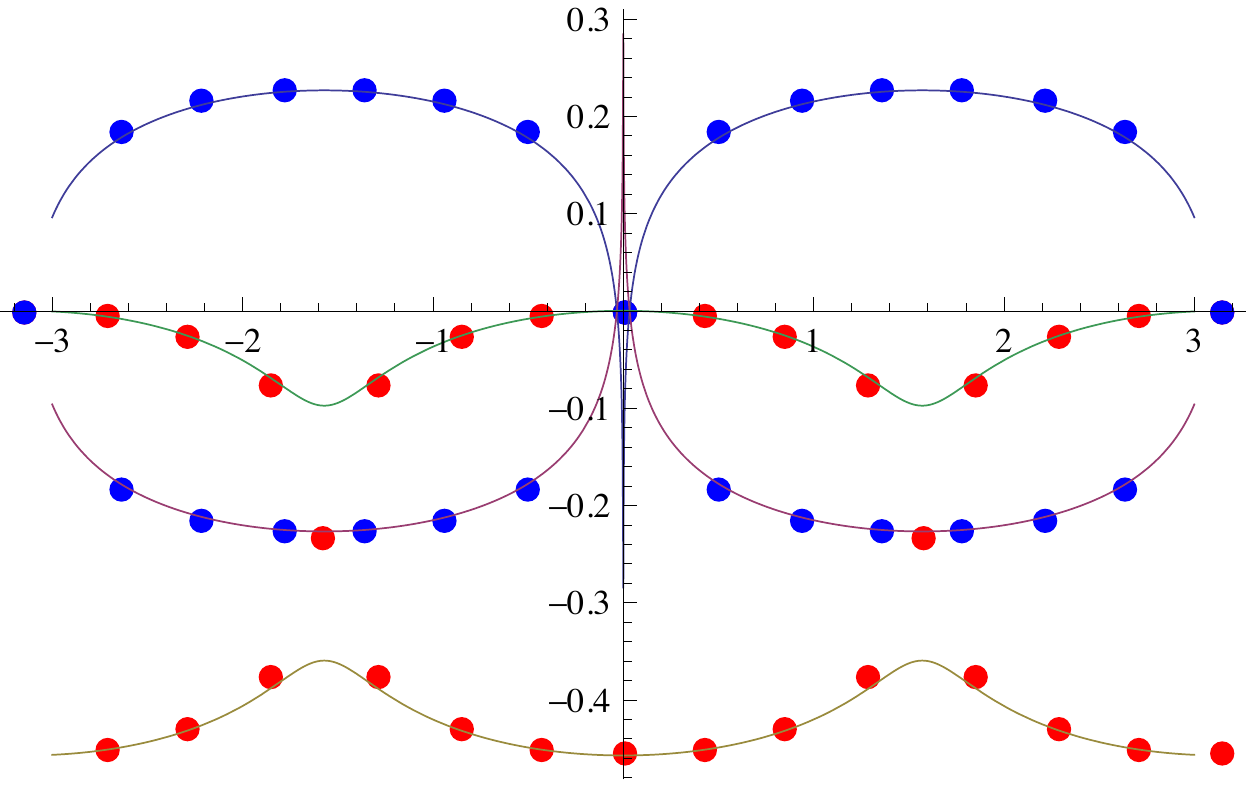} \hspace{1pt}
\includegraphics[width=5cm]{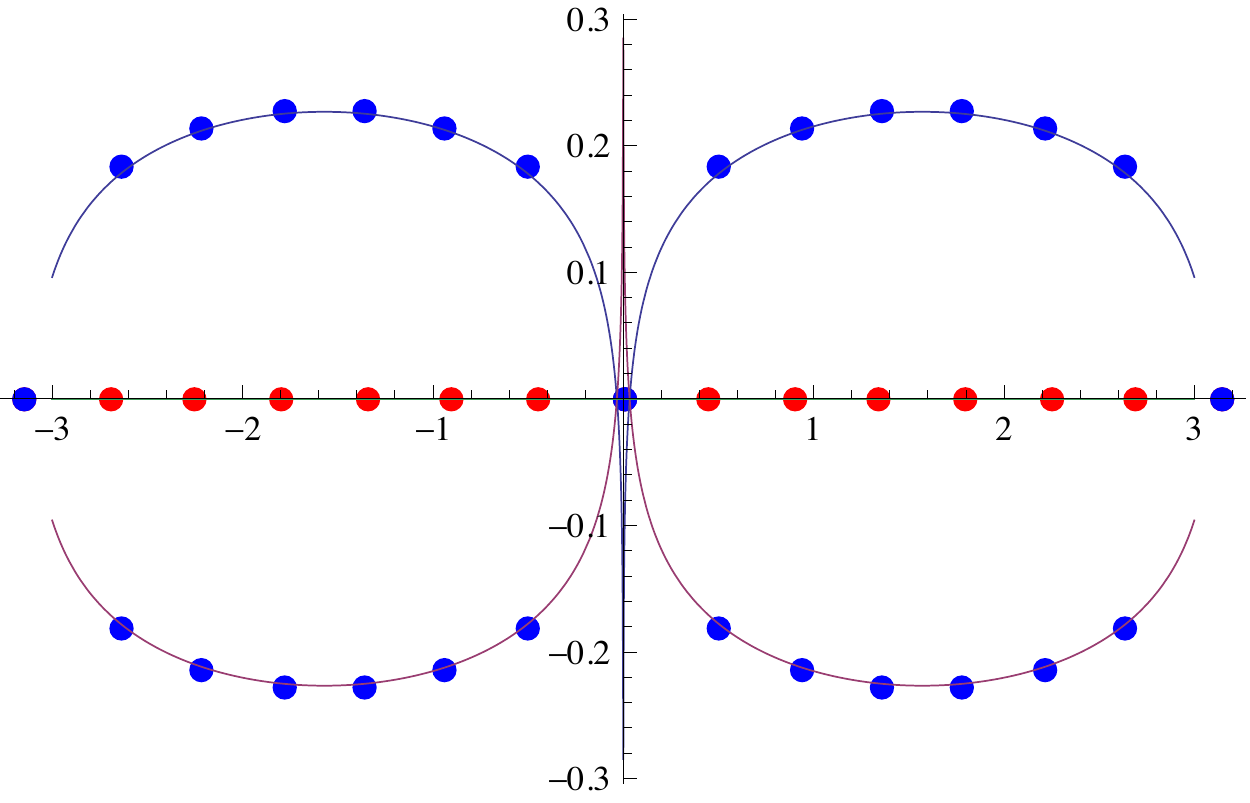}

 \caption{$\Sigma_{7,8,\kappa}$ with $\kappa=7.5$ (L), $\kappa=7.6$ (M),
$\kappa=8$ (critical, R).}
\label{WK9}
\end{figure}

Formula~\eqref{movingcurve} works well away from the critical values $\kappa=m$ or $\kappa=n$, when we have the
equilibria corresponding to $\psi=\sin nz/ W_{m,n}$ and $\psi=\sin mz/ W_{m,n}$ respectively (see Figs.~\ref{WK7} and~\ref{WK9}).
Conjecturally the formulae~\eqref{triglimcurve}, \eqref{movingcurve} def\/ine the asymptotic curves for the zeros in the
limit of large~$m$ and~$n$ with f\/ixed dif\/ference $n-m$ (cf.~\cite{FHV}, where the Wronskians of Hermite polynomials were studied).

\begin{figure}[h!] \centering \includegraphics[width=5cm]{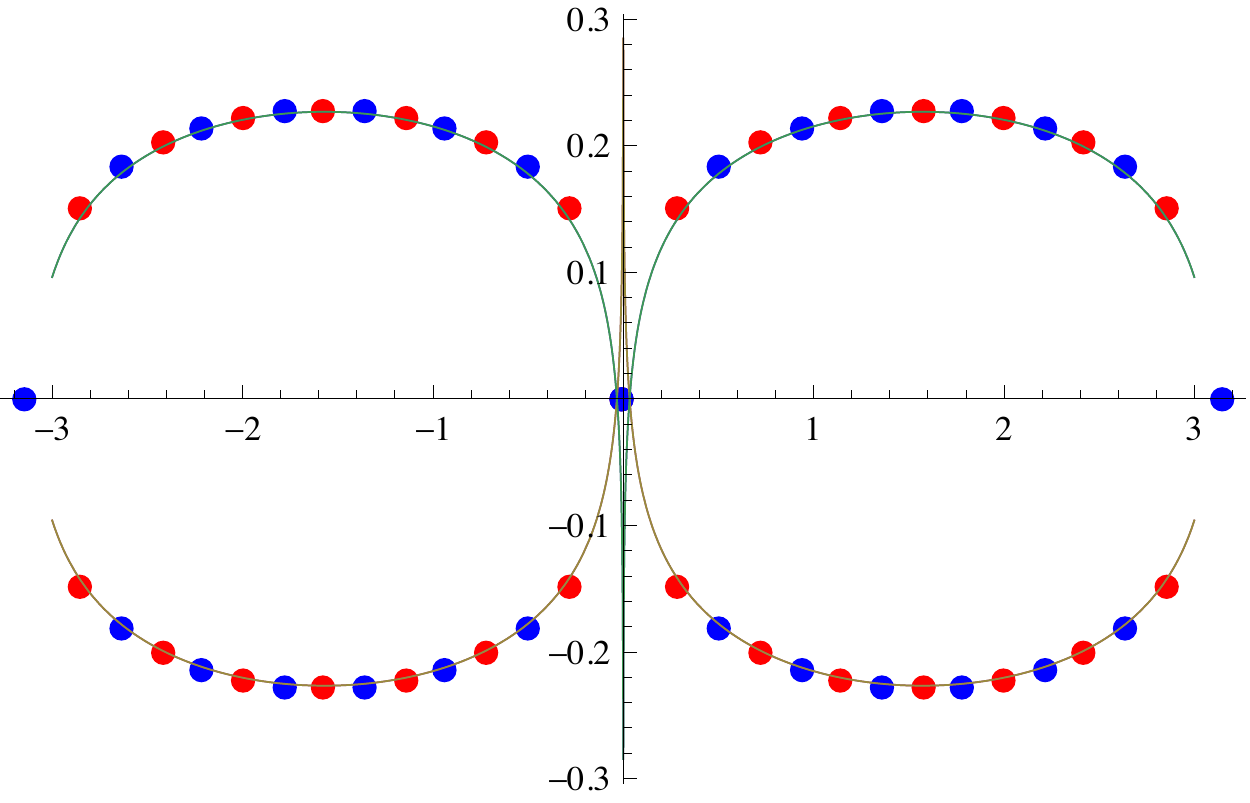} \hspace{1pt} \includegraphics[width=5cm]{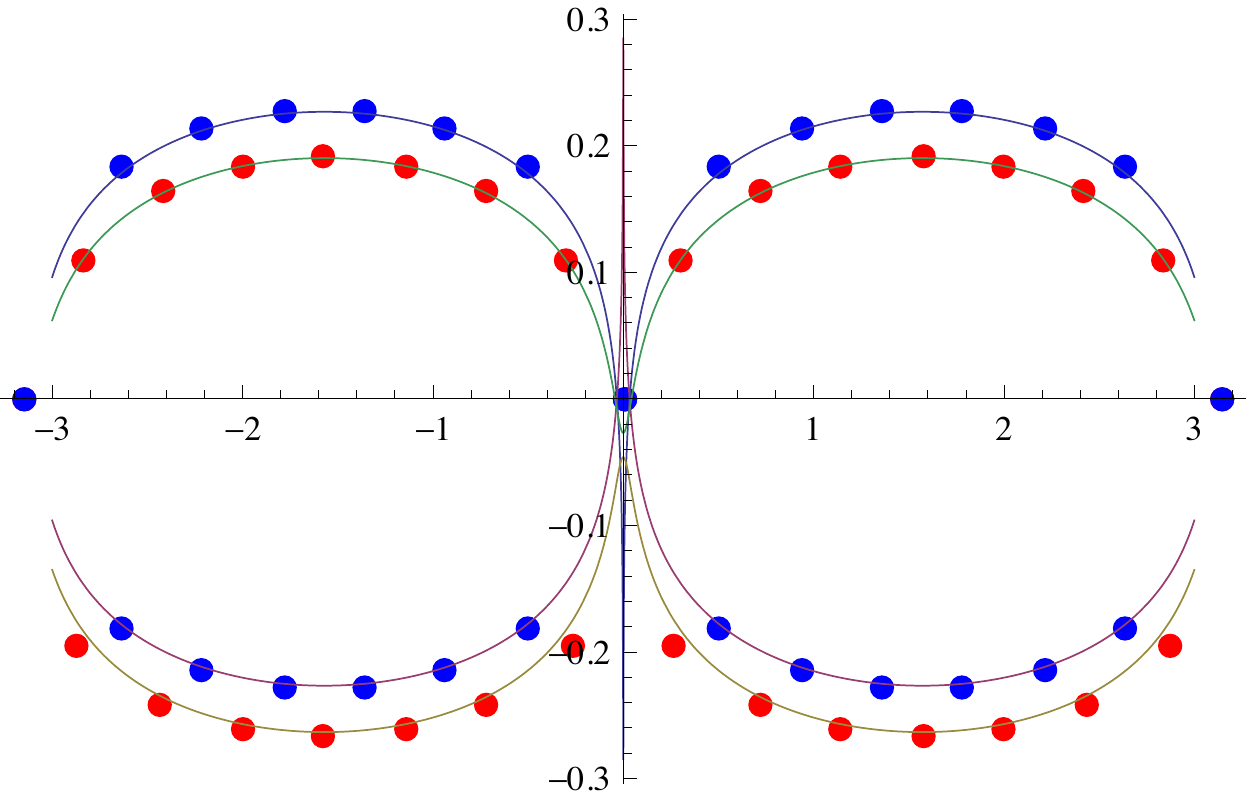} \hspace{1pt}
\includegraphics[width=5cm]{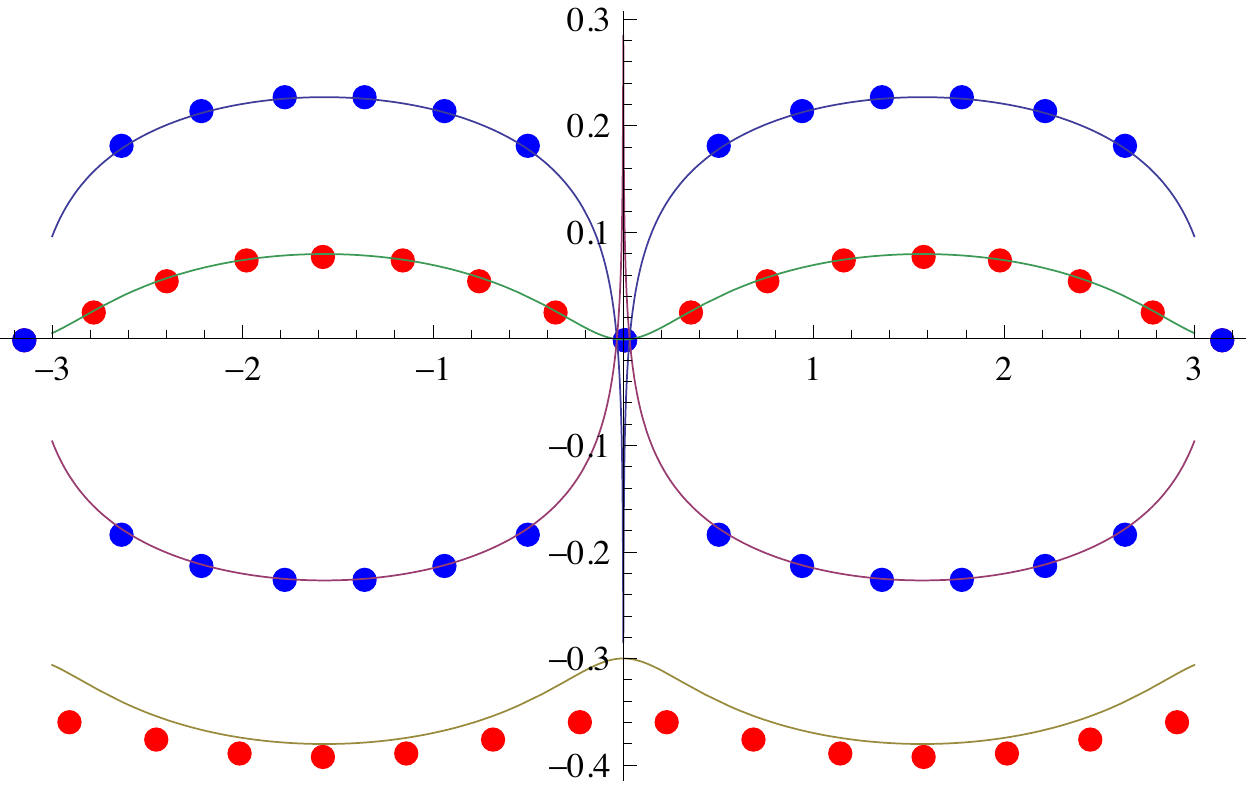}

 \caption{$\Sigma_{7,8,\kappa}$ with $\kappa=0$ (equilibrium, L), $\kappa=2$ (M),
$\kappa=6$ (R).}
\label{WK6}
\end{figure}

\begin{figure}[h!] \centering \includegraphics[width=5cm]{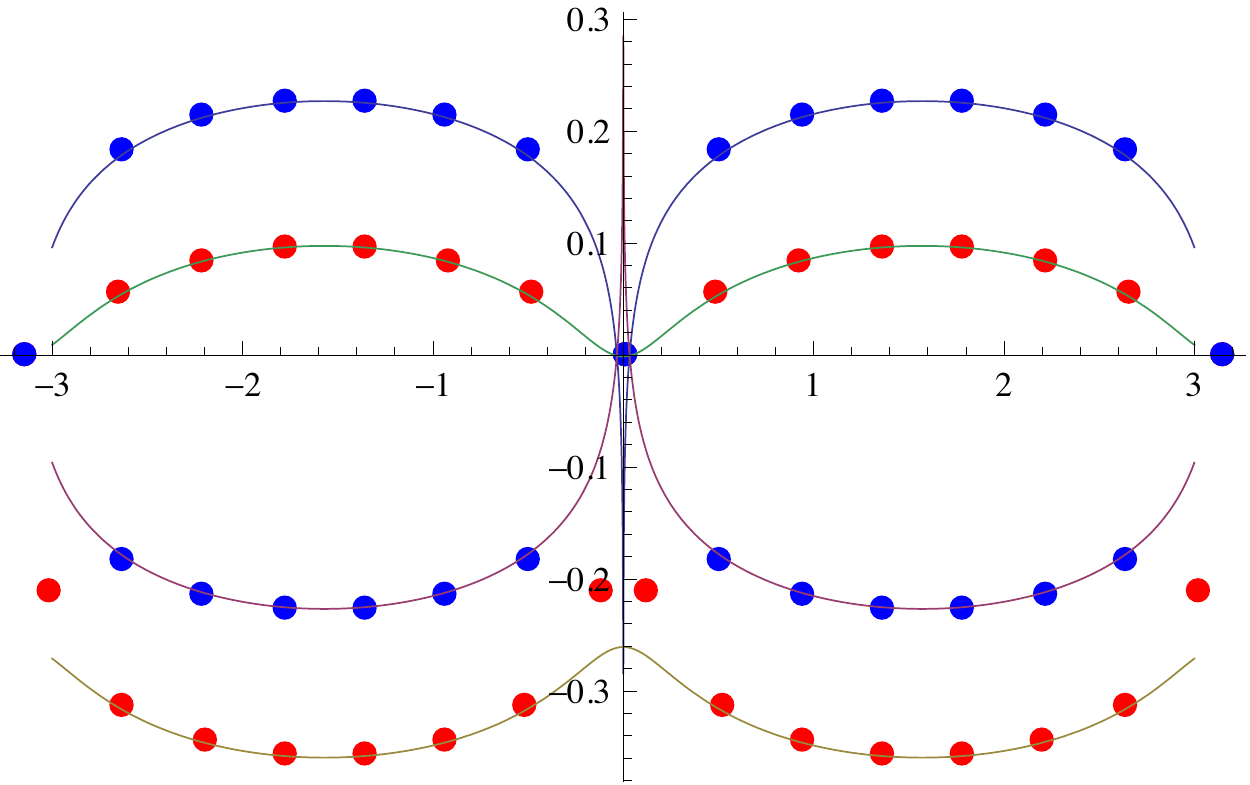} \hspace{1pt} \includegraphics[width=5cm]{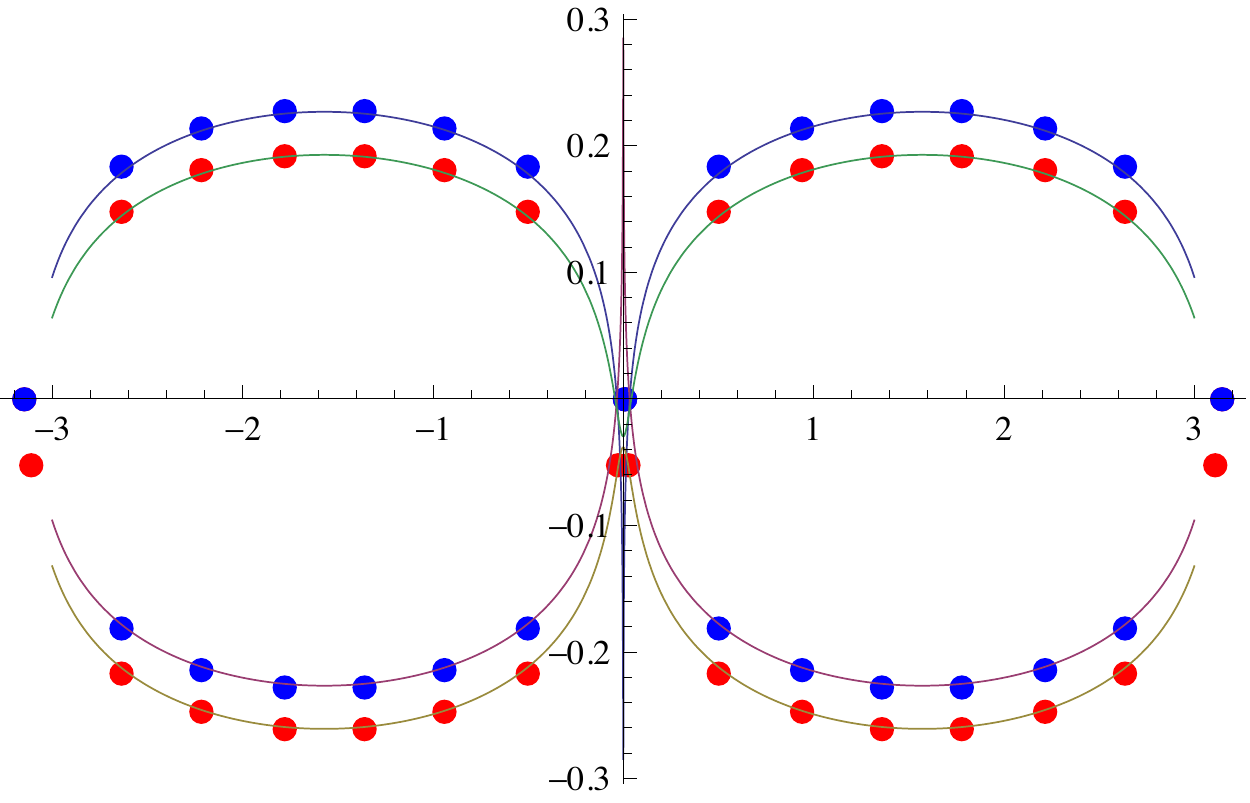} \hspace{1pt}
\includegraphics[width=5cm]{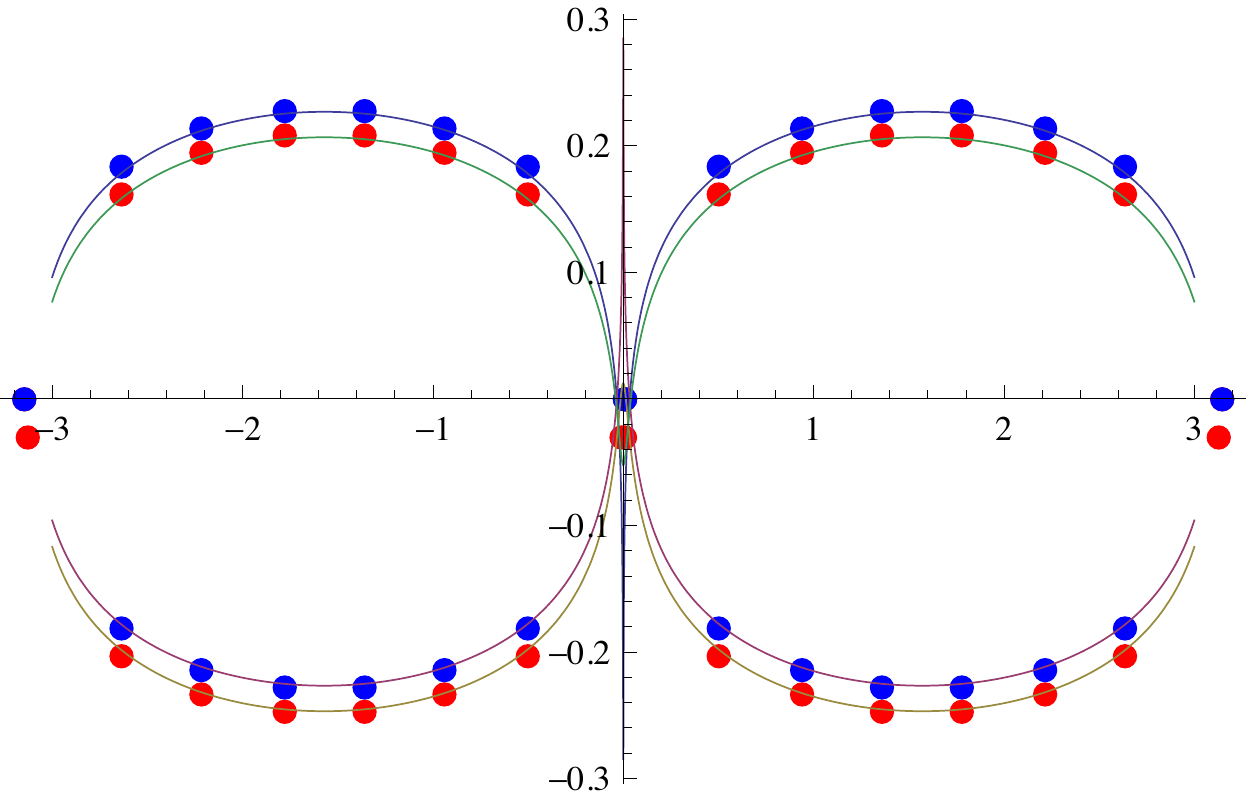}

 \caption{$\Sigma_{7,8,\kappa}$ with $\kappa=10$ (L), $\kappa=30$ (M),
$\kappa=50$ (R).}
\label{WK10}
\end{figure}

The conf\/igurations corresponding to $m=7$ and $n=8$ are displayed in the sequence of pictures shown in
Figs.~\ref{WK6}--\ref{WK10} for increasing values of the parameter~$\kappa$.
We give a~qualitative description of what is happening at each stage:
\begin{itemize}\itemsep=0pt
\item $\kappa=0$. The zeros of $W_{m,n}$ and $W_{m,n}^\kappa$ are interlaced.
\item $0<\kappa<m$. The zeros of $W_{m,n}^\kappa$ move downward whilst maintaining a~similar overall form until~$\kappa$
approaches~$m$, when they f\/latten out.
\item $\kappa=m$. The f\/irst critical value.
The bottom line of vortices tends to $-i\infty$.
The top line of vortices sit on the real axis at the zeros of $\sin nx$.
\item $m<\kappa<n$. The bottom line returns and a~vortex is exchanged between the top and bottom lines.
\item $\kappa=n$. The second critical case.
Again, the bottom line of vortices tends to $-i\infty$ and the top line of vortices sit on the real axis, only this time
at the zeros of $\sin mx$.
\item $\kappa>n$. The red vortices move upwards and tend towards the blue vortices as $\kappa\to\infty$.
\end{itemize}
Recall that in all these examples the velocity is horizontal since~$\kappa$ is real.


\section{Concluding remarks}

An explicit description of all monodromy-free operators is known only in a~few cases: rational class of potentials
decaying or with quadratic growth at inf\/inity~\cite{Obl} and trigonometric class described in Section~\ref{Section3}.
Already in the sextic rational case the situation is far from clear, see~\cite{GV} for the latest results in this
direction.
The same is true about monodromy-free perturbations of Whittaker--Hill operator~\cite{HV} and already mentioned elliptic
case.

The link with vortex dynamics adds one more reason to the importance of these problems.
It would be interesting to analyse from this point of view a~class of the monodromy-free potentials in terms of the
Painlev\'e-IV transcendents described in~\cite{V}.
Another interesting question is to study the quasi-periodic case by allowing in the construction of
$\Sigma_{k,\phi,\kappa}$ non-integer $k_j$.

The geometry of the trigonometric conf\/igurations $\Sigma_{k,\phi, \kappa}$ is also worthy of studying further, in
particular, the asymptotic shape of the corresponding vortex streets.
It would be nice also to see if the shape of the corresponding Young diagram of~$k$ plays any role here, similarly to
the case of Hermite polynomials in~\cite{FHV}.

Finally, from the point of view of possible applications the stability of the new equilibria is crucial and is to be
investigated (see Lamb~\cite{Lamb} for the conditions of stability for the original von K\'arm\'an streets).

\subsection*{Acknowledgements}

We are very grateful to John Gibbons and Boris Khesin for helpful and encouraging discussions.
We also thank all the referees for the critical comments and constructive suggestions.
This work was mainly done in spring 2012 when the f\/irst author (ADH) was a~PhD student at Loughborough University.
The work of APV was partly supported by the EPSRC (grant EP/J00488X/1).

\pdfbookmark[1]{References}{ref}
\LastPageEnding

\end{document}